# *TRAVAIL, FORCE VIVE ET FATIGUE DANS L'ŒUVRE DE DANIEL BERNOULLI : VERS L'OPTIMISATION DU FAIT BIOLOGIQUE (1738-1753)*

Yannick Fonteneau ; Jerome Viard

*Université Lyon 1 ; EA 4148 S2HEP*

*Résumé*

*Un antécédent du concept de travail mécanique réside dans les concepts de* potentia absoluta *et de travail des hommes, mis en œuvre dans la section IX de l'Hydrodynamica. Daniel Bernoulli ne confond pas ces deux entités : il existe un rapport de genre à espèce entre la première, générale, et la seconde, organique. Bernoulli distingue nettement force vive et* potentia absoluta *(ou travail) : leurs conversions mutuelles sont rarement mentionnées explicitement dans cet ouvrage de 1738 en dehors d'un exemple de conversion de force vive en travail dans la section X dans une problématique de substitutions des forces motrices et conditionnée à la médiation machinique. Son attitude évolue notablement dans un texte de 1753, où travail et force vive se trouvent explicitement connectés, tandis que le concept de* potentia absoluta *est réduit à celui de travail des hommes, et que le mot même est abandonné. Le travail peut alors se convertir en force vive, mais la réciproque n'est vraie que dans un seul cas, intra-organique. C'est le concept de fatigue, modélisée comme une dépense d'esprits animaux conçus eux-mêmes comme de petits ressorts bandés libérant de la force vive, qui permet la conversion, jamais quantifiée, de la force vive en travail. De la sorte, le travail peut apparaître* in fine *comme un état transitionnel entre deux formes de force vive, la première étant inquantifiable. Parallèlement, les éléments naturels sont discrédités de toute velléité de production rentable. Seuls les hommes et les animaux peuvent* travailler. *La nature, laissée à elle-même, ne travaille pas. Malgré sa volonté de rapprocher mécanique rationnelle et mécanique pratique, on perçoit chez Bernoulli la subsistance d'une disjonction rarement dépassée des domaines appliqués et théoriques.*

*Mots clés : Daniel Bernoulli, Travail mécanique, Force vive, Fatigue, Esprits animaux, Elasticité, Rendement, Optimisation, Galères, Rames.*

*Abstract*

*The concept of mechanical work is inherited from the concepts of* potentia absoluta *and men's work, both implemented in the Section IX of Daniel Bernoulli's Hydrodynamica in 1738. Nonetheless, Bernoulli did not confuse those two entities: he defined a link from gender to species between the former, general, and the latter, organic. Besides, Bernoulli clearly distinguished vis viva and* potentia absoluta *(or work). Their mutual conversions are rarely explicitly mentionned in this book, except once, in the Section X of his work, from* vis viva *to work, and subordinated to the mediation of a machine, in a driving forces substitution problem. His attitude significantly evolved in a text in 1753, in which work and* vis viva *were unambiguously connected, while the concept of* potentia absoluta *was reduced to the one of men's work, and the expression itself was abandoned. It was then accepted that work can be converted into* vis viva*, but the opposite is true in only one case, the intra-organic one. The concept of tiredness, seen as an expenditure of animal spirits conceived themselves as little tensed springs liberating* vis viva*, allowed direct conversion, even never quantified and listed simply as a model, from* vis viva *to work. Thus, work may have ultimately appeared as a transitional state between two kinds of* vis viva*, which the first is non-quantifiable. At the same time, natural elements were discredited from any hint of profitable production. Only men and animals were able to work in the strict sense of the word. Nature did not work by itself, according to Bernoulli. Despite his will to bring together rational mechanics and common mechanics, one perceived in the work of Bernoulli the subsistence of a rarely crossed disjunction between practical and theoretical fields.*

# INTRODUCTION

L'histoire du concept de travail mécanique se mêle intimement à celle de la constitution de la science des machines. De Guillaume Amontons, l'un des premiers à quantifier cette notion en 1699, jusqu'à Coriolis (1792-1843), qui sacre l'avènement officiel du concept de travail mécanique dans son ouvrage *Du Calcul de l'Effet des Machines*[1] en 1829, on assiste à une suite ininterrompue de tentatives visant à rapprocher les domaines de la science pragmatique et de la mécanique rationnelle au travers des antécédents de ce concept.[2] De ce point de vue, l'œuvre de Daniel Bernoulli est remarquable : loin de se contenter de créer ou de parfaire une simple mesure de l'effet des machines, il cherche véritablement à relier la notion de travail mécanique aux autres concepts qu'il utilise. Ainsi on ne saurait bien comprendre comment Bernoulli agit de ce point de vue sans se référer à sa conceptualisation de la force vive et de sa conservation, à son concept de force vive potentielle, à la fatigue, ainsi qu'au poids qu'a pour cet auteur une conception élastique de la matière, en directe influence de son père Jean Bernoulli et de Leibniz avant lui. On ne saurait le comprendre non plus sans prendre conscience du contexte qui motive ses démarches, l'homme au travail et la production de la machine, et ses catégories conséquentes de rentabilité et d'optimisation.

Mais quels sont ces antécédents du travail mécanique présents chez Daniel Bernoulli, et pourquoi peut-on les appeler ainsi ? Que recouvraient-ils exactement dans l'esprit de cet auteur ? Ces conceptions ont-elles évolué au cours de sa carrière ?

---

[1] G.-G. CORIOLIS, *Du calcul de l'effet des machines, ou considérations sur l'emploi des moteurs et sur leur évaluation, pour servir d'introduction à l'étude spéciale des machines*, Paris, Carilian-Golury, 1829.

[2] Précisons que Coriolis, célébré généralement comme l'inventeur du concept, œuvre en fait avec tout un groupe d'ingénieurs savants, qui intègre par exemple Navier, savants influencés également par Lazare Carnot et Coulomb. Poncelet, un peu plus tard, apportera également d'importantes contributions à ce concept.
Pour une histoire de la constitution du concept de travail mécanique depuis la fin du $17^e$ s. et dans le premier $18^e$ s., on pourra consulter : Y. FONTENEAU, *Développements précoces du concept de travail mécanique (fin 17e s.-début 18$^e$ s.) : quantification, optimisation et profit de l'effet des agents producteurs*, Thèse de doctorat, Lyon, Université Claude Bernard Lyon 1, 2011. Cf. également : Y. FONTENEAU, *Les antécédents du concept de travail mécanique chez Amontons, Parent et Daniel Bernoulli : de la qualité à la quantité (1699-1738)*, «Dix-Huitième Siècle», n° 41, 2009, pp. 343-368. Pour la période allant jusqu'au $19^e$ siècle, cf. D.S.L. CARDWELL, *Some factors in the early development of the concepts of Power, Work and Energy*, «The British journal for the history of science», 3, n° 11, 1967, pp. 209-224. Ainsi que : I. GRATTAN-GUINESS, *Work for the workers : Advances in engineering mechanics and instruction in France*, 1800-1830, «Annals of science», 41, 1984, pp. 1-33. Pour une histoire du concept au $19^e$ siècle mettant en avant l'influence des aspects économiques cf. K. CHATZIS, *Economies, machines et mécanique rationnelle: la naissance du concept de travail chez les ingénieurs-savants français, entre 1819 et 1829*, « Annales des Ponts et Chaussées », nouvelle série, n° 82, 1997, pp 10-20, et F. VATIN, *Le Travail, Economie et physique, 1780-1830*, Paris, P.U.F., 1993. Sur la période globale du $17^e$ s. au $19^e$ s., on trouve d'importants développements dans : J.-P. SERIS, *Machine et communication*, Paris, Vrin, 1987, notamment pp. 283-317, 343-376, 407-450.

Afin donc de saisir la véritable dimensionnalité des conceptions de Daniel Bernoulli, nous commencerons dans une première partie par montrer ce qu'elles ne sont pas, en contrepoint d'un article de Kevin C. de Berg, où l'auteur se permet de parler d'énergie cinétique à propos de la force vive, et d'identifier sans précautions ce que Bernoulli nomme *force vive potentielle* au travail mécanique.

Pour répondre à nos questions, nous procéderons ensuite chronologiquement, en étudiant les deux textes principaux où apparait cette notion : l'un d'entre eux, étudié dans la deuxième partie, étant sa célèbre *Hydrodynamica*,[3] et l'autre, traité dans la troisième partie, ayant remporté le prix de l'Académie Royale des Sciences de Paris de 1753.[4]

Dans ceux-ci, l'auteur met en place ce que nous dénommons un *antécédent* du concept de travail mécanique, à travers ses concepts de *potentia absoluta*, et de travail des hommes et des animaux, le premier étant, comme on le verra, plus large que le second. Sans tomber dans les critiques généralement évoquées à propos des précurseurs,[5] nous chercherons à comprendre quelle réalité cognitive particulière recouvrait les conceptions de Bernoulli, et à démanteler des argumentations abusives ayant traduit trop précipitamment les textes pour les interpréter en un sens par trop moderne. En effet, retracer l'histoire d'un concept est un exercice ardu, dont beaucoup de commentateurs ne se tirent qu'au prix d'une identification du concept avec son expression formelle. Méthode qui entraîne presque mécaniquement une projection de nos propres représentations sur celles des anciens.

Nous verrons en particulier comment le travail (dans ses acceptions de *potentia absoluta* ou de travail des animaux), bien que porté par une volonté de faire se rapprocher les domaines pratique et théorique, reste en grande partie cloisonné dans la science des machines, et ne réussit pas, entre les mains d'un Bernoulli pourtant autant « théoricien » que « pragmatique », à transcender sa matrice originelle pour s'imposer comme un concept central dans la Mécanique, ce qu'il ne fera que sous l'impulsion de Coriolis et de son groupe d'ingénieurs-savants, près de 90 ans après Bernoulli. A cette occasion nous verrons que le concept de *potentia absoluta* ne sera pas réutilisé chez cet auteur, quand bien même l'expression figure dans un texte mineur de 1768 traitant des frictions mais pour désigner tout à fait autre chose.[6]

---

[3] D. BERNOULLI, *Hydrodynamica, sive De viribus et motibus fluidorum commentarii. Opus academicum ab auctore, dum Petropoli ageret, congestum*, Strasbourg, Dulssecker, Decker, 1738. On pourra consulter plus facilement la récente édition commentée : D. BERNOULLI, *Die Werke von Daniel Bernoulli, Band 5, Hydrodynamik II*, Basel, Boston, Berlin, Birkhaüser, 2002. Il existe également une version anglaise plus ancienne : D. BERNOULLI, *Hydrodynamics*, Trad. de *Hydrodynamica, sive De viribus et motibus fluidorum commentarii (1738)* par T. CARMODY & H. KOBUS, New York, Dover Publications INC, 1968. Une traduction française a également été effectuée, mais nous ne la donnons ici que pour mémoire, sa qualité étant fort relative : D. BERNOULLI, *Hydrodynamique*, Trad. de *Hydrodynamica sive de viribus et motibus fluidorum commentarii* par J. PEYROUX, Bordeaux, J. Peyroux, 2004.
[4] D. BERNOULLI, *Recherches sur la manière la plus avantageuse de suppléer à l'action du vent sur les grands vaisseaux*, in *Recueil des pièces qui ont remportés le prix de l'académie royale des sciences*, vol. VII, Paris, Panckoucke, 1769 (1753). Cette édition comporte le désavantage de ne pas donner les figures. On pourra y palier grâce à l'édition complète et de grande qualité qui en a été faite par Frans CERULUS : D. BERNOULLI, *Die Werke von Daniel Bernoulli. Band 8, Technologie II*, éd. par F.A. Cerulus, Basel, Boston, Berlin, Birkhaüser, 2004, pp.184-251.
[5] Notamment dans l'épistémologie bachelardienne. Cf. G. CANGUILHEM, *L'objet de l'histoire des sciences*, in *Études d'histoire et de philosophie des sciences*, Paris, Vrin, 1968, pp. 9-23. G. BACHELARD, *L'Activité rationaliste de la physique contemporaine*, Paris, P.U.F., 1951. G. BACHELARD, *L'Actualité de l'histoire des sciences*, in *L'Engagement rationaliste*, Paris, P.U.F., 1972, pp. 137-152.

Nous verrons donc dans la seconde partie comment Bernoulli crée cet antécédent, la *potentia absoluta*, en exposant en détail ses caractéristiques, et comment il se place au sein d'une logique d'optimisation. Nous verrons ensuite dans la troisième partie comment ces conceptions évoluent le travail s'arcboute entre les concepts de fatigue et de force vive, comment il révèle la philosophie sociale de son auteur, et comment il s'articule avec sa conception élastique de la matière. Suivra la conclusion.

---

[6] D. BERNOULLI, 1987, *Commentation de utilissima ac commodissima directione potentiarum frictionibus mechanicis adhibendarum*, in D. BERNOULLI, *Die Werke von Daniel Bernoulli, Band 3, Mechanik*, éd. par D. Speiser, Basel, Boston, Stuttgart, Birkhäuser Verlag, pp. 209-218. (Publication originale : Novi commentarii Academiae Scientiarum Imperialis Petropolitanae, vol. XIII, 1769 (1768), pp. 242-256,).

# DES ILLUSIONS DE L' 'OPTIQUE FORMELLE'

Dans l'histoire de la physique, une époque prit fin, si l'on nous permet d'être schématique, autour de 1850, avec l'invention du principe de conservation de l'énergie. Celle-ci fut permise par la rencontre de deux traditions ; une tradition technique et pragmatique de calcul du travail des agents producteurs, et l'autre plus théorique de recherche d'un principe de conservation dans la nature sur la base des forces vives.[7]

Au début du siècle, le travail, jusqu'ici relativement isolé de la mécanique rationnelle, entre alors dans celle-ci grâce aux efforts d'un groupe d'ingénieurs savants parmi lesquels Navier, Poncelet, Coriolis joueront un rôle déterminant dans cette introduction. Coriolis[8] notamment, à la suite de Carnot en qui il reconnaît un précurseur selon Gillispsie[9], établit une relation d'équivalence entre ce concept et la force vive. Coriolis en soumettant la seconde au premier, met le travail au centre de sa nouvelle description[10]. Le travail mécanique apparaît de cette manière comme le passeur entre l'ancienne et la nouvelle physique, entre une physique de la force, et une physique de l'énergie.

Pour le résumer formellement, le travail mécanique s'exprime par l'intégrale $\int F \cdot dx$, où $F$ représente la force appliquée en un point, et $dx$ l'élément infinitésimal de longueur parcourue par le point d'application d'une force. Dans le cas le plus simple, une balle de poids P tombant en chute libre d'une hauteur H, le travail du poids[11] est égal à P.H. Nous nous contenterons pour l'instant de cet embryon de définition, mais, bien entendu, cette simple définition formelle ne saurait suffire à définir le concept dans sa globalité. Auquel cas, cette expression étant souvent présente dans des intermédiaires de calcul, nous pourrions en conclure abusivement que ce concept est présent partout depuis des temps immémoriaux. C'est précisément ce que semblent faire plusieurs auteurs, dont Kevin C. De Berg dans son article *The Development of the Concept of Work : A Case where History Can Inform Pedagogy*.[12] L'auteur de cet article y énonce sans ambages :

> La relation entre 'travail' et 'force vive' fut exprimée pour la première fois par Daniel Bernoulli en 1738 dans un article[13] où il considère la compression d'air dans un cylindre. Il appelle (poids x distance), (P+ p)x, la force vive potentielle, et ½ (P+ p)vv, la force vive réelle. Bien qu'il ne distingue pas la masse du poids, et n'utilise pas le terme 'travail', il est clair que sa dérivation mathématique est un important précurseur des concepts de travail, énergie potentielle et énergie cinétique.[14]

---

[7] Le fait de savoir si l'énergie doit plus à l'une de ces deux traditions qu'à l'autre a longtemps été source de querelles parmi les historiens. Il nous semble qu'elle se dissout quand on remarque que c'est leur convergence qui a permis cette invention. D'après Thomas Kuhn, la plupart des auteurs ayant pris part à cette nouveauté n'ont pas pris comme modèle la conservation des forces vives (à l'exception notable de S. Carnot, Mayer, et Helmholtz), mais leur source d'inspiration fut une tradition technique vieille de plus d'un siècle utilisant un concept de travail utilisé pour lui-même, et non mis en relation avec la force vive. (T.S. KUHN, *Un exemple de découverte simultanée : la conservation de l'énergie*, in T.S. KUHN, *La tension essentielle, Tradition et changement dans les sciences* (Trad. de l'anglais par Michel BIEZUNSKI…[et al.]), Paris, Gallimard, 1990, pp. 111-156 : pp. 132-133.

[8] G.-G. CORIOLIS, *Du calcul de l'effet des machines, ou considérations sur l'emploi des moteurs et sur leurs évaluation, pour servir d'introduction à l'étude spéciale des machines*, Paris, Carilian-Golury, 1829.

[9] C.C. GILLIPSIE, *Lazare Carnot, Savant*, Paris 1979 (Trad. fr., Princeton 1971) :p. 101, cité par J.-P. SERIS, *Machine et communication*, Paris, Vrin, 1987 : p. 346

[10] D'après O. DARRIGOL, 2001, *God, waterwheels, and molecules : Saint-Venant's anticipation of energy conservation*, « HSPS », 31, Part 2: 317, Poncelet ira « plus loin que Navier et Coriolis en fondant la formulation de la mécanique sur le concept de travail. »

[11] Nous verrons que cette expression de 'travail du poids' n'a, chez Daniel Bernoulli, aucun sens, les éléments naturels ne pouvant pas seuls réaliser un travail.

Ainsi donc d'après De Berg, Bernoulli non seulement inventerait le travail dans le texte cité, mais en plus, le relierait avec la force vive. Dans la suite, il se fait encore moins nuancé, puisqu'il énonce que la dérivation de Bernoulli est très instructive car il montre

> comment l'énergie cinétique (qu'il appelle 'force vive réelle' ) est reliée au travail produit (qu'il appelle 'force vive potentielle').[15]

De Berg, donc, croît que l'expression (P +p).x citée précédemment est un travail, et ½ (P+p).v.v de l'énergie cinétique. Il va même jusqu'à dire que l'on peut traduire le propos de l'auteur par l'expression moderne bien connue : *mgh= ½ m.v²* (p. 522), parlant ainsi de travail du poids. A propos du même texte, Pacey et Fisher faisaient les mêmes erreurs, trente ans plus tôt, en affirmant :

> Dans ce raisonnement, la descente du poids correspond à la dépense de travail et augmente une "vis viva potentielle" […]; mais la "vis viva potentielle" n'est pas clairement distinguée du travail, que Bernoulli appelle autre part "potentia absoluta"[16]

Par ailleurs, Brett D. Steele, relatant une application du calcul de la force vive aux projectiles de mousquets, qui suit immédiatement le paragraphe analysé par De Berg et Pacey & Fisher, ne remet pas en cause cette interprétation, tout juste pondérée par quelques guillemets :

---

[12] K.C. DE BERG, *The Development of the Concept of Work : A Case where History Can Inform Pedagogy* «Science & Education», n° 6 1997, pp. 511-527.
[13] Il s'agit en fait du §40 de la section X de son *Hydrodynamica*.
[14] Kevin C. DE BERG, *The Development of the Concept of Work : A Case where History Can Inform Pedagogy*, Science & Education 6, 1997. P. 514-515. Nous traduisons.
[15] *Ibidem*, p. 521-522.
[16] A.J. PACEY &S.J. FISHER, *Daniel Bernoulli and the vis viva of compressed air*, «The British journal for the history of science», 3, n° 4, 1967, pp. 388-392: 389: « In this reasoning, the descent of the weight corresponds to the expanditure of work and gives rise to a "potential vis viva" […] ; but "potential vis viva" is not clearly distinguished from work, which Bernoulli elsewhere calls "potentia absoluta" ». Nous traduisons. D'ailleurs dans le passage en question, Bernoulli n'utilise pas plus le mot de travail que le concept.

Bernoulli montra que le "travail" effectué par l'air expansé est égal à l' "énergie cinétique" du projectile à la bouche du fusil.[17]

Nous allons voir que tout ceci est erroné. En effet, cette traduction, ou plutôt cette bijection totale entre conceptions bernoullienne et moderne, est irrecevable pour l'historien des sciences, et ce pour plusieurs raisons, que nous allons d'abord mentionner puis développer par la suite.

La première est que l'essentiel de l'entreprise de Coriolis a consisté à réunir les concepts de force vive et de travail (et Coriolis lui-même est convaincu de la nouveauté de la chose), en arguant de leur conversion mutuelle, et à réinterpréter la physique à l'aune de ce dernier.[18] Si l'on suit De Berg, on ne comprend alors guère pourquoi ce n'est pas plutôt Daniel Bernoulli qui fut célébré comme l'inventeur du concept de travail mécanique, et on est conduit à conclure que Coriolis et ses collègues ingénieurs, formés à la meilleure école de l'époque, Polytechnique, étaient dans l'ignorance de l'œuvre majeure du 18$^e$ s. en matière d'hydraulique, ce qui semble pour le moins improbable.

La seconde est que Bernoulli ne s'autorise jamais à parler de *travail du poids*. Il s'interdit tout à fait, comme nous allons le voir, que les éléments naturels, livrés à eux-mêmes, sans médiation, puissent exercer un travail.

Troisièmement, le terme de « force vive potentielle » désigne quelque chose de bien précis dans l'esprit de Bernoulli, qui n'a rien à voir avec un travail. On pourrait croire, à tort, que le terme d'énergie potentielle serait plus approprié. Il n'en est rien, et nous verrons de quoi il retourne plus loin. Néanmoins, si De Berg se contentait de traduire la « force vive potentielle » de Bernoulli en « énergie potentielle », le malentendu serait moins grave : on comprendrait plus aisément qu'un pédagogue comme De Berg ait pu utiliser une traduction approximative, peu au fait des subtilités de la pensée de Bernoulli. Parler de travail, en revanche, ne correspond même pas à une traduction approximative de sa pensée. De Berg montre en outre qu'il fait une confusion, fréquente par ailleurs au sein des manuels scolaires, entre travail et variation d'énergie potentielle.

La quatrième raison, enfin, est que le concept d'énergie, en plein 18$^e$ s., n'avait pas émergé. Traduire ainsi « force vive » en « énergie cinétique », sans précaution, semble ainsi quelque peu cavalier. Nous verrons en effet que ce terme, chez Bernoulli, ne saurait être directement traduit de la sorte.

---

[17] B.D. STEELE, *Muskets and Pendulums: Benjamin Robins, Leonhard Euler, and the Ballistics Revolution*, «Technology and Culture», 35, n° 2, 1994, pp. 348-382: 358. « Bernoulli showed that the "work" done by expanding air is equal to the "kinetic energy" of the projectile at the muzzle. » Nous traduisons.
[18] Ce renversement conceptuel en faveur du travail mécanique figure déjà chez L. Carnot, dont on sait (J.-P. SERIS, *Machine et communication*, Paris, Vrin, 1987, p. 346-347) qu'il a été peu lu de son temps et redécouvert par la génération suivante dont fait partie précisément Coriolis. (L. Carnot, *Principes de l'équilibre et du mouvement*, Paris, Bachelier, 1803 : pp. 35-36, § 56-57 par exemple).

A présent, afin de corroborer ces quatre affirmations, il est nécessaire d'expliciter plus avant la démarche de Bernoulli. Regardons donc de plus près la section incriminée.

### .A. LA SECTION X : UNE PROBLEMATIQUE D'AIR COMPRIME

Dans la section en question, la dixième de l'Hydrodynamica,§40, donc, Bernoulli considère un piston ABCD rempli d'air atmosphérique, situé dans un vide d'air (Voir Figure 1).

La distance EB est posée égale à *a*. La partie EF est mobile et soutient tout d'abord un poids *p*, « égal à la pression »[19] qu'exerce l'air atmosphérique lorsque le piston est dans des conditions normales. Dans celles-ci, puisque l'air intérieur est de l'air atmosphérique, il peut soutenir le poids *p* qu'on lui impose. Ainsi EF est en équilibre. A présent, posons un poids supplémentaire *P* par-dessus le poids *p*. Le plateau EF va alors se

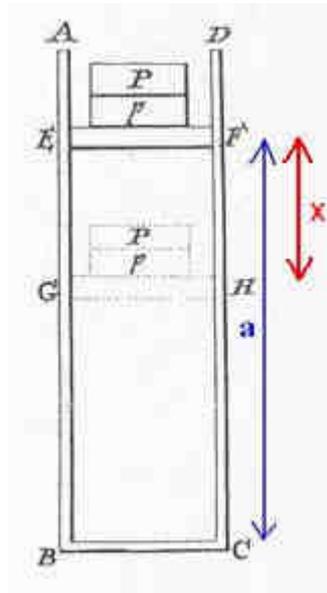

Figure 1 : Piston dans un vide d'air, rempli d'air pouvant soutenir le poids *p* à la hauteur *a*.

mouvoir vers le bas, avec une certaine vitesse. Supposons qu'à une certaine distance *x* de EF le plateau ait la vitesse *v*. grâce à la loi de Boyle-Mariotte (le produit de la pression, assimilé à un poids, par le volume, reste constant, si la température est constante), Bernoulli peut calculer la pression, dirigée vers le haut, de l'air intérieur comprimé qui va s'exercer sur EF. On trouve alors facilement que :

P1.V1=P2.V2, avec les indices 1 pour les pressions initiales et 2 pour les finales, donc $P_2 = \frac{P_1 \cdot V_1}{V_2} = \frac{p \cdot (a \cdot \Pi b^2)}{[(a-x) \cdot \Pi b^2]}$ avec b le rayon de la base du cylindre, d'où : $P_2 = \frac{p \cdot a}{(a-x)}$

D'où la force résultante sur le plateau EF égale à : $P + p - \frac{a}{a-x} \cdot p$

---

[19] Le terme de pression, ici, ne désigne pas une *pression* au sens moderne, mais ce que nous appellerions une *force de pression*, c'est-à-dire le produit de la pression par une surface, qu'il peut égaliser à un poids.

Utilisant alors le concept newtonien de force accélératrice, Bernoulli va diviser cette force résultante par la masse de l'objet qui se meut, ce qui lui donne la « force accélératrice » (donc $dv/dt$), et multiplier le tout par la différentielle de temps $dt$, ou ce qui est la même chose, $dx/v$, pour obtenir la différentielle de la vitesse $dv$. Donc :

$$dv = \frac{\left(P + p - \dfrac{ap}{a-x}\right) \cdot \dfrac{dx}{v}}{(P+p)}$$

Ayons soin de remarquer dans cette expression que le P+p du numérateur est une force, newtonienne, tandis que celui du dénominateur est une masse. Il ne reste plus qu'à faire passer P+p dans le premier membre et à intégrer. D'où :

$$\frac{1}{2}(P+p)vv = (P+p)x - ap \cdot \ln\frac{a}{a-x}$$

C'est cette formule qui permet de dire à De Berg que si la compression de l'air n'opposait pas de résistance, alors on obtiendrait ½ (P+p).v² = (P+p)x , c'est-à-dire la première relation explicite entre la force vive et le travail, ce que De Berg traduit par : mgh = ½ mv².

Mais que nous dit Bernoulli ? Il appelle la force vive effectivement acquise par le poids (P+p) lors de sa chute, depuis la hauteur x, contrariée par la pression de l'air, *force vive réelle* (« *vis viva actuali* »[20]), c'est à dire ½ (P+p) v² (le membre de gauche de la dernière équation). Face à cela, nous avons la *force vive potentielle* (« *vis viva potentialis* »), c'est à dire la force vive acquise par le poids (P+p) si celui-ci tombait en chute libre depuis la hauteur x. Cette force vive est *potentielle* puisqu'elle exprime ce qui pourrait se passer dans des conditions qui ne sont pas celles ci-dessus évoquées, ou, si l'on veut, *la force vive totale susceptible d'être disponible au cours de la chute*. Cette force vive potentielle est donc égale à ½ (P+p)V² (avec ici V désignant non plus le volume, mais la vitesse qu'acquérait le poids (P+p) si celui-ci tombait en chute libre depuis la hauteur x), ou bien, en vertu du principe de conservation des forces vives, à (P+p)x . La différence entre cette dernière expression et la force vive réelle est donc donnée par la seconde partie du membre de droite dans l'équation ci-dessus. L'interprétation physique que Bernoulli en donne correspond à la quantité de force vive qui a été transmise à l'air pour que celui-ci se comprime. Et comme cette quantité est exprimée par le produit du poids p et d'un nombre, Bernoulli peut énoncer que la quantité de force vive nécessaire à comprimer un certain volume d'air jusqu'à une certaine valeur, est égale à celle qu'acquerrait ce même poids p tombant d'une hauteur de [a.ln a/(a-x)] pieds.

---

[20] D. BERNOULLI, *Die Werke von Daniel Bernoulli. Band 8, Technologie II*, éd. par F.A. Cerulus, Basel, Boston, Berlin, Birkhaüser, 2004: 344 (Sect. X, §40).

Pour parler concrètement, supposons que la surface de EF soit d'un pied carré, et que le poids p soit égal à 2240 livres, c'est-à-dire précisément le poids correspondant à la pression de l'air atmosphérique exercée sur une surface d'un pied carré. Si a=2 pieds, et qu'on veut comprimer notre volume d'air de manière à lui donner une densité deux fois plus élevée, il faut donc que x soit égal à 1 pied. En pratiquant cette compression, nous utilisons donc une force vive de :

$$2 \cdot \ln \frac{2}{2-1} \cdot 2240 = 3105$$ (Bernoulli ne donne pas d'unités),

c'est-à-dire la force vive acquise par un corps de masse p'=3105 livres tombant d'une hauteur de 1 pied, ou , ce qui est la même chose, d'un corps de 1552.5 livres tombant de 2 pieds, et ainsi de suite de tous les produits donnant 3105. Selon l'axiome de Huygens de l'égalité de la descente réelle et de la remontée potentielle,[21] Bernoulli peut alors énoncer réciproquement que si l'on dispose d'un air deux fois plus dense que l'air atmosphérique, celui-ci pourra élever un poids de 3105 livres à un pied de hauteur. Il calculera ensuite selon les mêmes méthodes un piston situé dans l'air et non plus dans le vide.

### .B. PAS DE TRAVAIL MECANIQUE, MAIS UNE PRENOTION D'ENERGIE INTERNE

Bernoulli expose-t-il donc ici une quelconque notion ou antécédent de travail mécanique ? Non. Nous venons de voir que le seul concept disponible dans ces lignes est celui de force vive, et que toute son argumentation se révèle n'être que *partie intégrante du schème de conservation de la force vive leibnizienne*. Le corps pesant ne travaille pas : il ne fait que générer de la force vive. La descente du corps fournit constamment la même quantité de force vive à celui-ci : nous ne faisons qu'assister, dans le mouvement de va-et-vient du piston, à la restitution de la force vive emmagasinée par l'air, transférée de celui-ci au piston et de ce dernier au premier. En ce sens le piston joue le même rôle que le pendule de Huygens, expérience-paradigme de l'égalité de la descente réelle et de la remontée potentielle. La même quantité de force vive qui avait servi à descendre est réutilisée pour remonter. A cette différence près cependant, que le mouvement du piston n'est pas autoentretenu. Il faut un agent extérieur si donc nous voulons, du moins dans ces conditions, faire en sorte que le piston s'abaisse et se relève, ce en quoi cette situation est très proche de celle du ressort. Il existe alors dans l'esprit de Bernoulli une similitude entre sa manière de concevoir un air comprimé, et la manière dont on pouvait concevoir un ressort[22].

---

[21] D. BERNOULLI, *Die Werke von Daniel Bernoulli, Band 5, Hydrodynamik II*, éd. par G.K. Mikhailov, Basel, Boston, Berlin, Birkhaüser, 2002: 108, § 18. Bernoulli parle au sujet de la conservation de la force vive d'
« *égalité entre la descente en acte et l'ascension en puissance* ». Selon J.-P. SERIS, *Machine et communication*, Paris, Vrin, 1987 : p. 302, ce principe constitue pour Bernoulli une « Hypothèse physique » fondamentale validée par ses conséquences « ses a poteriori » .

Ce qui est à l'œuvre ici donc n'a rien à voir avec une prénotion de travail, mais relève de tout autre chose. En effet la manière dont Bernoulli considère la force vive nous paraît novatrice en ce sens que celle-ci est emmagasinée dans l'air après sa transmission, et c'est la capacité de stockage de la force vive qui permet cette transmission.

Il faut aller plus loin dans le texte, et lire le § 43, pour qu'apparaisse timidement une notion de travail. Bernoulli, dès lors, change de registre : après avoir mis en place jusqu'ici un outil théorique, il en vient à traiter brièvement de l'air expansé par le moyen du charbon, comme dans les machines à feu. Il bascule dès lors dans le domaine pratique des machines productives, et va faire référence au travail des hommes. En effet, après avoir cité un mémoire de Guillaume Amontons de 1699, où ce dernier propose de tirer du travail de la puissance du feu grâce à un moulin la transformant en force productive,[23] il énonce une conviction :

Je suis convaincu que si toute la force vive *latente* d'un pied cube de charbon [vis viva, quae in carbonum pede cubico latet], tirée par le moyen du feu, était convenablement appliquée à la conduite d'une *machine*, on en tirerait plus de profit que du labeur d'une journée [labore diurno] de huit ou dix hommes.[24]

Une force vive latente ? Si Bernoulli ne créé pas dans cette section de notion de travail, il semble pour autant avoir l'intuition aiguë de ce qu'on appellera plus tard non pas tant une énergie potentielle qu'une énergie interne.[25] C'est donc de la sorte que Bernoulli

---

[22] Les propriétés des fluides élastiques sont développées par Bernoulli au §39 de cette même section.
[23] G. AMONTONS, *Moyen de substituer commodément l'action du feu à la force des hommes et des chevaux pour mouvoir les machines*, in *MARS 1699*, Paris, Martin, Coignard fils, Guerin, 1732, pp. 112-126 (Mémoires). Sur ce texte cf. Y. FONTENEAU, *Développements précoces du concept de travail mécanique (fin 17e s.-début 18e s.) : quantification, optimisation et profit de l'effet des agents producteurs*, Thèse de doctorat, Lyon, Université Claude Bernard Lyon 1, 2011, pp. 143-201. Quelques indications dans J.-P. SERIS, Machine et communication, Paris, Vrin, 1987, pp. 193-198.
[24] D. BERNOULLI, *Die Werke von Daniel Bernoulli, Band 5, Hydrodynamik II*, éd. par G.K. Mikhailov, Basel, Boston, Berlin, Birkhaüser, 2002: 346 (Sect. X, § 43). Nous soulignons. Nous traduisons.
[25] Nous reviendrons plus bas (§ F. de la troisième partie de cet article) sur ces notions de force vive potentielle et latente, et les liens qu'elles entretiennent avec la conception élastique de la matière chez Daniel et Jean Bernoulli. Si certains auteurs des premiers traités de thermodynamique au 19[e] s. ont eu tendance à confondre les notions d'énergie interne et de travail mécanique, D. Bernoulli ici différencie nettement force vive latente et travail des hommes. En outre, cette distinction est en accord avec les définitions prises par L. Carnot et Coriolis de la force vive comme équivalente à du travail potentiel. L'origine de la notion de force vive latente est attribuée à tort par certains auteurs du 19[e] s. à Lazare Carnot : « Le terme d'énergie est dû à Young, celui d'énergie potentielle semble avoir été émis pour la première fois par L.N.M. Carnot qui parle de force vive latente (Principes fondamentaux de l'équilibre et du mouvement 1803 ) et par W. Thomson , qui l'appelle énergie statique. » (P.-G. Tait, *Esquisse historique de la théorie dynamique de la chaleur* (Traduit par F. Moigno), Paris, Gauthier-Villard, 1870 : 72). Tait semble ignorer que le terme est déjà présent chez Daniel Bernoulli, comme nous le voyons dans ce paragraphe, bien que son emploi par Carnot soit réalisé dans un contexte légèrement différent : « Nous venons de voir que la force vive peut se présenter sous la forme Mu2

semble interpréter le dispositif d'Amontons : un formidable moyen de tirer la force vive latente du combustible pour la conduite d'une machine, dont l'effet remplacerait celui du labeur de bien des hommes. Il réitère de la sorte quelques lignes plus loin, toujours dans le même paragraphe, lorsqu'il calcule de la même manière la force vive consacrée à l'expansion d'un volume d'air ayant subi la déflagration d'une charge de poudre à canon, en concluant :

> Ainsi donc, en théorie, il existe une *machine* au moyen de laquelle un pied cube de poudre à canon peut élever 183 913 864 livres à la hauteur d'un pied, lequel travail [laborem], je le crois, même 100 hommes très forts ne pourraient pas accomplir en une journée, quelle que soit la machine qui est utilisée.[26]

Il faut noter que ce sont les deux seuls passages de la section X, d'ailleurs fort courts, et se limitant à une comparaison, où apparaît une notion de travail. Il semble alors que par ces quelques lignes, Daniel Bernoulli indique que la force vive va pouvoir être convertie en travail mécanique par le jeu d'un dispositif machinique (et seulement ainsi). La machine, mettant en acte la potentialité de la poudre à canon ou du charbon, va produire le mouvement ascendant d'un certain poids, mouvement tout à fait équivalent à celui obtenu par le travail d'un homme ou d'un animal, lequel travail est précisément défini par l'élévation d'un objet pesant à une certaine hauteur. Bernoulli se place dans la même logique que Guillaume Amontons en 1699, dont il semble par ailleurs très bien connaître les travaux,[27] c'est-à-dire une logique de substitution des forces mouvantes, ici la force des hommes et la force de l'air expansé, dont on peut comparer les effets.

Pour conclure, au terme de l'analyse de cette section, il faut insister sur l'influence du contexte sur la pensée de Bernoulli. Dans cette section essentiellement théorique, le contexte est ici celui de la conservation des forces vives aussi bien dans le domaine de la chute des corps que dans celui des corps élastiques. Le dispositif du §. 40 a pour but, de l'aveu de Bernoulli, d' « *établir une correspondance entre la conservation des forces vives contenues dans de l'air comprimé et un corps qui est tombé d'une certaine hauteur* » (§.43). Mais « *on ne peut espérer aucun avantage du dispositif précédent […] pour améliorer l'usage des machines* » (*ibidem*). Ce dispositif n'a qu'un objectif : fournir un moyen de calcul. Une fois

---

d'une masse par le carré d'une vitesse, ou sous la forme PH d'une force motrice par une ligne. Dans le premier cas, c'est la force vive proprement dite ; dans le second on pourrait lui donner la dénomination particulière de force vive latente » (L. Carnot, *Principes fondamentaux de l'équilibre et du mouvement*, Paris, Bachelier, 1803 : 37, §59). Plus largement, sur la construction de la notion d'énergie potentielle, on consultera avec profit la première partie de : S. Faye, *Elaboration et expérimentation d'une situation d'enseignement et d'apprentissage pouvant permettre à des élèves de première S du Sénégal de construire le concept d'énergie potentielle*, Thèse de Doctorat, Lyon, Université Claude Bernard Lyon 1, 2007.
[26] D. BERNOULLI, *Die Werke von Daniel Bernoulli, Band 5, Hydrodynamik II*, ed. par G.K. Mikhailov, Basel, Boston, Berlin, Birkhaüser, 2002: 347 (Sect. X, § 43). Nous soulignons. Nous traduisons.
[27] Dans cette même dixième section, Bernoulli cite en longueur un mémoire d'Amontons de 1702 traitant de l'élasticité de l'air, et en discute les conclusions (§. 6).

établi le calcul de la force vive contenue dans un fluide élastique, il est possible de passer à une application relative aux machines mues par l'action du feu telle que celle d'Amontons. C'est alors que Bernoulli en vient à comparer le travail susceptible d'être obtenu par une machine mue par la force du feu à celui produit par une machine mue par la force des hommes.

Si Bernoulli finit donc par mettre en rapport force vive et travail, c'est donc d'une manière qui n'a rien à voir avec la suggestion de De Berg. En outre, dans cette section, ce rapport est univoque, en ce que la conversion de travail en force vive n'est pas mentionnée. La conversion réciproque de force vive en travail n'est quant à elle évoquée que lorsque l'auteur en vient à traiter de problèmes pratiques. Enfin, Bernoulli n'évoque jamais une notion de *travail du poids* : la nature en général, et la pesanteur en particulier, ne sont pas douées de la capacité de produire du travail par *eux-mêmes*, au contraire des animaux, hommes compris. Il faudra toujours un intermédiaire, une machine, pour que la force vive, potentielle ou actuelle, puisse se convertir en travail.

Pour mieux comprendre ce que Bernoulli entend par travail, il va nous falloir revenir en arrière dans son *Hydrodynamica*, et nous intéresser au concept de *potentia absolua* qu'il met en place dans la section IX, dans un contexte ayant trait à la constitution d'une science des machines hydrauliques, où existent des déperditions et des frictions, et où l'effet d'une machine pourra être compris comme catégorie générale regroupant sous son nom effets utiles et effets inutiles. Mais qu'est-ce exactement que la *potentia absoluta* ?

# LA *POTENTIA ABSOLUTA*, UN ANTECEDENT DU TRAVAIL MECANIQUE

Ce terme de *potentia absoluta*[28] figure dans la section IX, dont le titre est déjà révélateur :

> Du mouvement des fluides jaillissant sous l'effet non de leur propre poids, mais d'une puissance *extérieure*, et plus particulièrement des machines hydrauliques et du degré ultime de perfection qu'elles peuvent atteindre, ainsi que du moyen de parfaire cela à l'avenir grâce à la mécanique des solides aussi bien que des fluides.[29]

Notons que dans les sections précédentes, Bernoulli n'avait traité que du mouvement de fluides mus par leur propre poids, et c'est cette condition qui lui permettait d'appliquer l'énoncé de Huygens de l'égalité de la descente réelle et de la remontée potentielle. Il procède alors à un saut qualitatif en traitant de la sorte certes toujours le mouvement, mais provoqué par un agent étranger aux fluides mêmes. Dans ce titre, tout est déjà dit, ou presque : s'intéresser aux puissances extérieures qui font mouvoir les fluides revient de fait à s'intéresser à des dispositifs qui vont diriger les fluides dans un but précis. Il s'agit en somme de considérer la sphère des applications pratiques. En outre, le fait de penser la machine sous un rapport de plus grande perfection, ouvre la voie à une problématique d'optimisation et de rentabilité typique de la sphère productive.[30] Enfin, la volonté de les parfaire par le biais de la mécanique rationnelle rend manifeste une tentative de dissoudre la distinction entre théorie et pratique, ce qui semble être l'une des ambitions de la science des machines, et une des raisons d'être du concept de travail mécanique.

## .A. UN NOUVEAU CONCEPT ADAPTE A LA SPHERE PRODUCTIVE

Bernoulli va-t-il alors utiliser les mêmes outils que dans les autres sections ? Non, la nature des sujets étant, dans cet exposé des choses, trop différente pour donner lieu à un traitement similaire. Il lui faut donc inventer de nouveaux outils. Les « définitions » du début de la section sont l'occasion de les exposer :

> J'entends par puissance mouvante [potentia movente] ce principe actif consistant en un poids, une pression en action, ou une autre force morte de ce genre.[31]

---

[28] Jean-Pierre Séris a consacré un important développement à ce concept *in* : J.-P. SÉRIS, *Machine et communication*, Paris, Vrin, 1987 : pp. 299-318. Ce terme a aussi été repéré par G.K.Mikhaïlov, qui y avait vu un concept de travail mécanique, mais sans s'en expliquer véritablement : G.K. MIKHAILOV, *Introduction to Daniel Bernoulli's Hydrodynamica*, in D. BERNOULLI, *Die werke von Daniel Bernoulli, band 5, Hydrodynamik II*, éd. par G.K. Mikhailov, Basel, Boston, Berlin, Birkhaüser, 2002, pp. 17-78: 61.

[29] D. BERNOULLI, *Die Werke von Daniel Bernoulli, Band 5, Hydrodynamik II*, éd. par G.K. Mikhailov, Basel, Boston, Berlin, Birkhaüser, 2002: 274. « De motu fluidorum, quae non proprio pondere, sed potentia aliena ajiciuntur, ubi praesertim de Machinis Hydraulicis aerundemque ultimo qui dari potest perfectionis gradu, & quomodo mechanica tam solidorum quam fluidorum ulterius perfeci posit ». Citation traduite par nous, ainsi que les suivantes.

[30] Une voie initiée par Antoine Parent 34 ans auparavant. Cf. A. PARENT, *Sur la plus grande perfection possible des machines*, in *MARS 1704*, Paris, Martin, Coignard, Guerin, 1745, pp. 323-338.

C'est donc cette puissance mouvante, inspirée de la force morte leibnizienne, et dimensionnellement exprimable par un poids, qui constitue cet agent externe forçant notre fluide à se mouvoir. Bref elle est la *cause* du mouvement. Le concept ne semble guère original mais Bernoulli se devait de le nommer. La suite est bien plus originale, lorsque Bernoulli présente l'effet de cette cause agissante :

> De plus, le produit qui vient de la multiplication de cette *puissance mouvante* par sa vitesse et également par le temps pendant lequel elle exerce sa pression, je le désigne par *puissance absolue* [potentia absoluta]. Ou, puisque le produit de la vitesse et du temps est simplement proportionnel à la distance couverte, il sera également permis de comprendre la *puissance absolue* comme la *puissance mouvante* multipliée par la distance dont *celle-ci* se meut.[32]

Mais quelle signification physique ce concept possède-t-il? La réponse est immédiate :

> J'appelle ce produit *puissance absolue* [potentia absoluta] car c'est à partir de lui que doit être estimé le travail enduré par les ouvriers [labores hominum operariorum] pour l'élévation des eaux, ce qui, je le montrerai bientôt, sera prouvé par les règles que je donnerai en cette matière.[33]

La première de ces règles s'énonce alors ainsi :

> Le travail des ouvriers appliqués aux machines hydrauliques pour l'élévation des eaux, doit être estimé par leur *potentia absoluta*, c'est-à-dire par la *puissance mouvante* ou pression qu'ils exercent, par le temps, et par la vitesse du point auquel la *puissance mouvante* est appliquée.[34]

Le travail des hommes sert donc ici de référence, et qui plus est, Bernoulli s'attache à montrer que celui-ci est bien représenté par la *potentia absoluta*. Pour ce faire, il donne, à la suite de cette première règle, trois preuves :

    a. Le labeur des hommes est directement proportionnel au nombre de travailleurs appliqués à l'ouvrage, et donc, *proportionnel à la puissance mouvante appliquée*, si on raisonne à vitesse constante et sur la même durée.

    b. Concernant le temps, si on l'augmente alors on augmente le labeur dans la même proportion

---

c. Enfin, que l'on double la puissance mouvante ou qu'on en double la vitesse, il se produit le même effet, c'est-à-dire, par exemple, que l'on élèvera la même quantité d'eau dans le même temps, au final.

Le poids, la vitesse et le temps, multipliés entre eux, sont donc trois paramètres pertinents pour l'évaluation du labeur des hommes, car ceux-ci se rapportent à l'effet produit. Ce produit étant la définition de la *potentia absoluta*, on peut conclure que le travail des hommes et une image directe de cette dernière.[35]

Soulignons immédiatement un point. Si Daniel Bernoulli utilise deux expressions ici, *potentia absoluta* et travail des hommes, c'est qu'il existe une distinction entre elles, ou plus exactement un rapport de genre à espèce: le travail des hommes est une forme de *potentia absoluta*, mais toutes les *potentia absoluta* ne proviennent pas de lui. En effet, ce qui importe dans la définition qu'il donne de son nouveau concept, est la référence à la *puissance mouvante*. C'est donc la pression résultante que l'on se doit de considérer, ainsi que la vitesse de cette pression et le temps pendant lequel elle agit. On voit immédiatement que dans le cas d'un homme, la pression qu'il exerce, en régime stationnaire, ne varie pas avec la vitesse, et qu'il existe une identité stricte entre sa vitesse propre et la vitesse du point d'application de son effort (une rame par exemple).

Il en va tout autrement d'une machine mue par un fluide tel que l'eau, comme dans le cas d'un moulin à eau.[36] Dans ce second cas, en effet, la pale fuira devant le fluide en mouvement, et pour connaître la force avec laquelle ce dernier frappera la pale, il faudra considérer la vitesse résultant de la soustraction de celle de la pale à celle de l'eau. Ainsi la pression exercée sur la pale mouvante par le fluide ne sera ni identique ni directement proportionnelle à la pression exercée sur une pale immobile. En conséquence, on ne saurait en toute rigueur parler de *potentia absoluta* d'un fluide, car les paramètres en jeu ne sont pas la pression totale du fluide (c'est-à-dire celle s'exerçant sur une pale immobile) ni la vitesse totale du fluide, mais en réalité la *puissance mouvante* et la vitesse de la pale : la *potentia absoluta* nait au contact avec les machines mues et caractérise l'effet exercé à l'entrée. C'est bien cette notion de mobilité qui change tout, et que Bernoulli souligne en parlant de puissance *mouvante* : il s'agit de considérer les forces en tant qu'elles sont capables d'une action continue, renouvelée et en mouvement. Ainsi donc, s'il existe une identité entre travail des hommes et *potentia absoluta*, cette dernière se distingue du premier dans son rapport de généralité, pouvant définir un effet exercé sur une machine résultant de n'importe quelle *puissance mouvante* donnée. Si le travail met directement en rapport la force de l'homme avec son effet exercé sur la machine, les éléments, par nature, ne peuvent exercer sur la machine leur force totale. La médiation de la machine, signe à la fois la condition de la production par les éléments, et leur nécessaire déperdition.

---

[35] D. Bernoulli ne prend donc pas la puissance mouvante comme mesure de l'effet des agents producteurs. Pourtant les auteurs de l'environnement français de l'époque n'ont pas encore clairement tranché quel définition de l'effet était la plus souhaitable : effet de la force à l'équilibre, ou effet de la force en mouvement. Carnot en 1803 se fait l'écho de ce débat : « il se présente deux manières aussi naturelles l'une que l'autre, d'évaluer l'action qu'ils exercent effectivement. L'une consiste à voir quel fardeau un homme, par exemple, peut porter, ou quel effort évalué en poids il peut soutenir, tout en demeurant en repos. […] on [l'] appelle quelquefois force morte. […] La seconde méthode […] est d'examiner l'ouvrage qu'il est en état de faire dans un temps donné […] par un travail suivi. […] Quand on emploie des ouvriers, l'intérêt est de savoir ce qu'ils peuvent faire de travail dans un genre analogue à celui dont on vient de parler, bien plus que de savoir les fardeaux qu'ils pourroient porter sans bouger de place. » (L. CARNOT, *Principes de l'équilibre et du mouvement*, Paris, Bachelier, 1803 : pp. 34-35, §55-56). Sur ce passage de Carnot, cf. J.-P. SERIS, Machine et communication, Paris, Vrin, 1987 : p. 364).
[36] Sur l'histoire et la théorie des roues hydrauliques, on consultera : T.S. REYNOLDS, *Stronger than a hundred men : a history of the vertical water wheel*, Baltimore, Johns Hopkins University Press, 1983. Voir aussi B. BELHOSTE and J.-F. BELHOSTE, *La théorie des machines et les roues hydrauliques*, « Cahiers d'histoire et de philosophie des sciences »: 29 (1990), 1-17.

Nous voyons donc un terme, la *potentia absoluta*, dont les dimensions sont identiques à celles du travail mécanique, pris dans sa formulation moderne, acquérir le statut de concept autonome, et, qui plus est, indexé sur le labeur, le travail, des hommes. C'est par ce concept que doit être mesuré l'effet appliqué à la machine, en considérant la puissance mouvante résultante, et en référence au travail des hommes. Celui-ci va constituer le socle de ses réflexions durant toute la première partie de la section IX, et il va bien le différencier du facteur essentiel et limitant que constitue la fatigue.

## .B. LA FATIGUE, UN PARAMETRE PROBLEMATIQUE MAIS DETERMINANT DANS LA CONCEPTION DES MACHINES

Le poids, la vitesse et le temps, multipliés entre eux, sont donc trois paramètres pertinents pour l'évaluation du labeur des hommes, car ceux-ci se rapportent à l'effet produit. Mais ce ne saurait être une mesure de la fatigue. Bernoulli le précise immédiatement :

La précédente proposition ne doit pas être interprété dans un sens physiologique mais dans un sens moral[37]

Qu'est-ce à dire? Les différentes manières d'obtenir le produit *P.v.t* sont moralement[38] identiques, car elles produisent des effets identiques mais elles ne sont pas équivalentes d'un point de vue physiologique :

---

[37] *Ibidem*: 276 (§ 4). « Propositio praecedens non sensu physiologico sed morali est interpretanda ». Nous traduisons.

[…] moralement [moraliter] je n'attribue ni plus ni moins de valeur au travail d'un homme qui exerce à la même vitesse un effort double qu'à celui qui faisant le même effort double la vitesse, parce que certainement l'un ou l'autre produisent le même effet, quoiqu'il puisse arriver que le travail de l'un bien qu'il ne soit pas moins fort que l'autre soit beaucoup plus grand en un sens physiologique.[39]

Parmi les différents facteurs qui interviennent dans la mesure du travail la vitesse a une place à part si l'on se place d'un point de vue physiologique. Ainsi, nous dit-il (p. 276) si quelqu'un parcoure avec un effort de 20 livres une distance 200 pieds dans la première minute, il pourra aisément doubler l'effort (passant ainsi à 40 livres), mais il pourra beaucoup plus difficilement doubler sa vitesse (pour atteindre donc 400 pieds par minute). Bien que dans les deux cas la *potentia absoluta* résultante soit double, la fatigue physiologique éprouvée par le deuxième homme sera bien plus grande. C'est pourquoi Bernoulli prend soin de faire une remarque importante :

il faut considérer particulièrement comment chaque type de machine doit être constitué de sorte que pour la fatigue minimale des hommes [minima hominum defatigatione] le produit de leurs efforts par les vitesses soit en même temps un maximum[40]

Il convient donc, nous dit Bernoulli, de prendre en compte la fatigue dans la conception des machines car cela permet d'adapter les machines aux dispositions naturelles de l'organisme humain.[41] En effet derrière la distinction entre fatigue et travail (ou *potentia absoluta*) c'est la distinction entre 'puissance' [42] et travail qui se cache. Doubler la vitesse avec laquelle une tâche est réalisée revient à doubler la 'puissance' nécessaire à l'exécution de la tâche. Or la puissance moyenne disponible dépend directement de la constitution de l'organisme. Si l'on ne respecte pas les données physiologiques, la fatigue va être beaucoup plus grande pour un même travail. Mais comment prendre en compte la fatigue si on ne peut pas la mesurer, puisqu'elle n'est pas identique ou même proportionnelle au travail fourni ? En faisant référence à l'expérience. Et Bernoulli prend un exemple celui des cages d'écureuil. Marcher dans une cage d'écureuil équivaut à effectuer une marche sur un plan incliné. Or un marcheur pourra faire l'ascension d'une hauteur de plusieurs milliers de pieds pendant une dizaine d'heures, ce qui correspond à la durée d'une journée de travail dans des conditions de fatigue minimale si l'inclinaison du chemin parcouru est correctement choisie : ni trop forte, ni trop faible. Et il arrive à la conclusion que l'angle optimal est un angle de 30° ce qui correspond pour le marcheur à élever la moitié de son poids pendant sa marche. Bernoulli en conclut que l'écartement des marches des cages d'écureuil doivent correspondre précisément pour l'homme qui est à l'intérieur à une marche sur un plan faisant une inclinaison de 30° avec l'horizontale.

---

[38] Cet adjectif de moral fait peut-être référence à Descartes et à la certitude morale, celle qui suffit dans la pratique, au quotidien, et trouve sa place dans la construction du concept de probabilité. Par exemple, je sais *moralement* que Rome existe, même si je n'y suis jamais allé, car j'ai rencontré de nombreuses fois des personnes ayant été en cette ville, mais je n'en suis pas *absolument* sûr. Il semble de même que, chez Bernoulli, les différentes manières d'obtenir le produit P.v.t sont moralement identiques, car elles génèrent des fatigues identiques : en effet, dans l'immense majorité des cas, on ne rencontre que des travaux qui demandent à être exécutés pendant de longue période de temps (toute une journée), donc qui génèrent des fatigues raisonnables et à peu près égales. Mais dans des cas extrêmes (grande vitesse, ou forte charge), il n'en va plus de même, et deux manières de former le même produit P.v.t généreront des fatigues différentes. Les différentes manières de former P.v.t ne sont donc pas *absolument* identiques, au sens où P.v.t ne représente pas *absolument* la fatigue. Cf. R. DESCARTES, *Les principes de la philosophie, IV*, in V. COUSIN (ed.), *Œuvres de Descartes*, vol. III, Paris, F.G. Levrault, 1824 (1644), pp. 526-[1]. Cf. également A. ARNAULD &P. NICOLE, *La logique, ou L'art de penser : contenant, outre les regles communes, plusieurs observations nouvelles propres à former le jugement*, Paris, Charles Savreux, 1662: IV, 13. On trouve également l'occurrence chez J. BERNOULLI, *Ars conjectandi, opus posthumum. Accedit Tractatus de seriebus infinitis et epistola gallice scripta de ludo pilae reticularis*, Basileae, Thurnisiorum fratrum, 1713: p. 226. Cet écrit a été réédité in : J. BERNOULLI, *Die Werke von Jakob Bernoulli*, vol. 3, Basel, Birkhaüser, 1975. Leibniz également utilise cette notion. En ce qui concerne la littérature secondaire, on consultera à profit I. HACKING, *L'émergence de la probabilité*, Trad. de *The Emergence of probability* par M. DUFOUR, Paris, Seuil, 2002.
[39] D. BERNOULLI, *Die Werke von Daniel Bernoulli, Band 5, Hydrodynamik II*, éd. par G.K. Mikhailov, Basel, Boston, Berlin, Birkhaüser, 2002: 276. Nous traduisons.
[40] *Ibidem*: 276 (§ 4). Nous traduisons.
[41] Cf. J.-P. SERIS, *Machine et communication*, Paris, Vrin, 1987 : p. 307 :« Labeur n'est pas fatigue, mais son estimation juste est la condition d'une étude physiologique de la fatigue et d'une rationalisation du travail humain. »
[42] Puissance est prise ici au sens actuel en physique d'une énergie fournie en un temps donné

Ces conceptions se retrouveront quasiment mot à mot un demi-siècle plus tard dans le discours d'un Coulomb, contributeur de l'entrée du concept de travail dans la physique théorique au 19$^e$ siècle,[43] et également lecteur de Bernoulli :

> […] en supposant que nous ayons une formule qui représente l'effet, et une autre qui représente la fatigue, il faut, pour tirer le plus grand parti des forces animales, que l'effet divisé par la fatigue soit un maximum.[44]

La fatigue est donc un facteur problématique pour Bernoulli ici au sens où il ne peut la calculer via la *potentia absoluta*. Il faut donc prendre garde que les machines soient réglées de sorte à ce qu'elles provoquent la fatigue minimale pour un travail donné. Ainsi, un même travail provoquera une même fatigue, et c'est uniquement à cette condition que Bernoulli pourra raisonner à partir de son concept de *potentia absoluta*, dont l'utilité est de caractériser le travail humain. Si chaque opérateur fonctionne à fatigue minimale pour un travail donné, on sera alors en mesure de comparer différents travaux entre eux. Bernoulli s'attarde peu sur cette question de la fatigue, mais il nous apprend tout même que sa dépendance est fonction des mêmes trois paramètres que la *potentia absoluta*, quoique d'une manière plus composée, sans pouvoir préciser la nature de cette composition. La vitesse, notamment, semble jouer plus fortement sur la fatigue résultante. Nous verrons que dans un texte de 1753, dont l'analyse figure en dernière partie de cet article, il prendra la peine de justifier dans le détail ses positions.

---

[43] F. VATIN, *Le Travail, Economie et physique, 1780-1830*, Paris, P.U.F., 1993: 36-56. Chez Coulomb, le terme de quantité d'action remplacera celui de puissance absolue.
[44] C.-A. COULOMB, *Résultat de plusieurs expériences Destinées à déterminer la quantité d'action que les hommes peuvent fournir par leur travail journalier, suivant les différentes manières dont ils emploient leurs forces*, in BACHELIER (ed.), *Théorie des Machines Simples en ayant égard au frottement de leurs parties et à la roideur des cordages*, Paris, 1821: 256. Ce texte est identique à celui figurant dans les mémoires de l'académie des sciences de 1799. Des versions antérieures avaient été présentées en 1778, 1780 et 1798. Il faut noter que, dans ce texte, Coulomb fait référence au texte de Bernoulli ayant remporté le prix de l'Académie Royale des Sciences de 1753, que nous étudions dans la dernière partie de cet article.
L. Carnot abordera également le problème de la physiologie, mais sans citer explicitement la fatigue, en disant « que chaque agent a, eu égard à sa nature ou constitution physique, un *maximum* [qui] ne peut en général se trouver que par expérience » (L. CARNOT, Principes de l'équilibre et du mouvement, Paris, Bachelier, 1803 : p. 252, § 276).

Le labeur d'un homme, son travail, est donc ce qu'il produit et non sa fatigue, même si celle-ci doit être prise en compte dans la conception des machines. Quelle que soit la manière dont l'homme emploie ses forces, ce qui est utilisé mécaniquement est l'*effet exercé* sur la machine par la *puissance mouvante*. La *potentia absoluta* ainsi comprise est l'image directe du travail des hommes (travail appliqué et non travail physiologique), mais, lorsque des éléments sont en jeu dans le mouvement mécanique, elle ne dit rien sur la force originelle des éléments, en dehors de leur médiation par la machine, et donc ne dit rien d'un quelconque 'travail' des éléments. En revanche, l'effet exercé sur la machine par un homme, alias la *potentia absoluta*, alias le labeur, le travail d'un homme, est quantitativement différente de la fatigue éprouvée par lui. Bernoulli choisit donc ici de mettre en avant le travail des hommes : c'est ici la valeur de référence, vraisemblablement car elle est perçue comme économiquement significative. C'est le travail des hommes que l'on paye, ou que l'on cherche à remplacer. C'est donc par rapport à lui que doit se fonder une échelle de comparaison, et c'est en grande partie dans cet état d'esprit que sont nées les premières notions de travail mécanique.[45]

.C. EFFET EXERCE, EFFET PRODUIT : LA MESURE DE L'IDEALITE.

---

[45] Amontons, en 1699 déjà, ne procédait pas autrement. Ainsi en est-il également de toutes les tentatives de comparer la force des hommes et des chevaux, ou leurs travaux respectifs (La première mention que nous connaissons d'une tentative expérimentale de ce genre de comparaison figure dans : ACADEMIE ROYALE DES SCIENCES &B.L.B.D. FONTENELLE, *Mémoires de l'Académie royale des sciences depuis 1666 jusqu'en 1699*, 11 vols, vol. 1, Paris, Compagnie des libraires, 1729: 70 sq.)

Mais l'effet *exercé sur* la machine est-il le même que l'effet *produit par* la machine ?

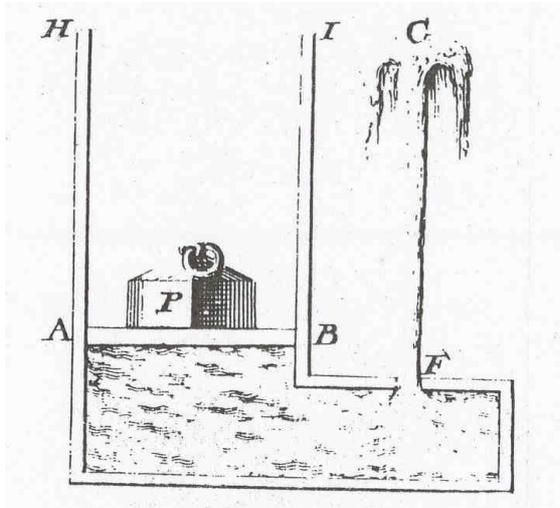

Figure 2 : Pompe la plus simple envisagée par Bernoulli

Non, à moins que nous faisions en sorte qu'il n'y ait nulle friction et nul effet inutile, comme stipulé dans la règle 2[46]:

> Avec la même *puissance absolue* donnée, je dis que toutes les machines qui ne souffrent d'aucune friction et qui ne génèrent aucun mouvement inutile pour la fin proposée maintiennent le même effet, et qu'on ne doit donc pas en préférer une à une autre. [47]

Nous comprenons bien la référence aux frictions, mais en quoi consistent exactement les mouvements inutiles susnommés? Pour bien le saisir, envisageons deux cas: la Figure 2 (correspondant à la figure 45 de la p. 274 dans le texte de Bernoulli) et la Figure 3 (numérotée 48 chez l'auteur, p. 279). Sur la première est représentée la pompe la plus simple que Bernoulli utilise. Le poids P est égal au poids du volume d'eau contenu dans la colonne HABI. Ainsi tout se passe comme si le niveau de l'eau montait jusqu'en HI et donc, de par le principe de l'égalité de la descente réelle et de la remontée potentielle, le jet sortant en F va monter jusqu'à la hauteur de HI, c'est-à-dire en G, si nous annulons mentalement les pertes, qui pourtant ne manqueraient pas de se produire sur les parois. Toute la puissance absolue est donc utilisée pour amener l'eau en G. Maintenant examinons la figure suivante : la partie gauche est identique, et le poids P correspond au poids du volume d'eau situé entre le niveau du plateau qui soutient le poids et le niveau se trouvant à la hauteur de G. Mais sur la partie droite la remontée de l'eau s'effectue dans un tuyau à partir de l'ouverture D, et qui s'incurve pour aboutir à la sortie F. Donc ici, l'eau s'est vue imposer une certaine puissance absolue qui aurait pu avoir comme résultat de faire remonter l'eau en G, mais ce résultat n'a pu avoir lieu, contraint par la structure à sortir alors qu'elle pouvait encore exercer un certain effet. On aurait donc pu avoir un poids P inférieur pour le même résultat, le même effet produit. On a donc gaspillé de la *potentia absoluta*, c'est-à-dire qu'on en a imprimé trop par rapport à ce qu'on voulait obtenir. De même, si l'on reprend la première figure, si nous nous arrangeons pour récupérer toute l'eau qui arrive en G, et qui a donc une vitesse nulle, nous pourrons dire que toute la *potentia absoluta* aura été utilisée. Mais si nous plaçons le bac disons un mètre plus bas, par exemple, alors l'eau va s'élever jusqu'en G puis retomber de 1 mètre, et ainsi nous aurons utilisé trop de *potentia absoluta* que nous n'en aurions dû. Voilà ce que Bernoulli appelle mouvement inutile. La meilleure machine, précise-t-il, est donc celle qui en génère le moins.[48]

---

[46] Dans cette régle, pour J.P. Séris, en introduisant une machine idéale : « Daniel Bernoulli ouvre ici la voie à Lazare Carnot (et à son fils) in J.-P. SERIS, *Machine et communication*, Paris, Vrin, 1987 : p. 309.
[47] D. BERNOULLI, *Die Werke von Daniel Bernoulli. Band 8, Technologie II*, éd. par F.A. Cerulus, Basel, Boston, Berlin, Birkhaüser, 2004: 277 (Sect. IX, § 5).Nous traduisons.

Ainsi donc dans ce cas seul, le cas idéal, l'effet exercé sur la machine, la *potentia absoluta*, et l'effet produit par la machine, en vienne à se confondre dans une symbiose fictionnelle. Car c'est bien une fiction, une fiction rationnelle, qui permet au phénomène général et au comportement réel de coïncider. C'est par différence vis-à-vis de cette chimère théorique atteinte par pure et arbitraire annulation des contraintes réelles signant l'impossibilité du mouvement perpétuel, sur cette représentation industrielle de l'idée de conservation, que va se penser le mouvement des machines réelles. Remarquons donc que la *potentia absoluta* est une mesure de ce que l'on applique à l'entrée, de l'*input*, et non pas de ce que l'on a à la sortie, de l'*output*, concept pour lequel Bernoulli utilise le terme d'effet, à moins bien sûr que l'on soit dans le cas idéal.

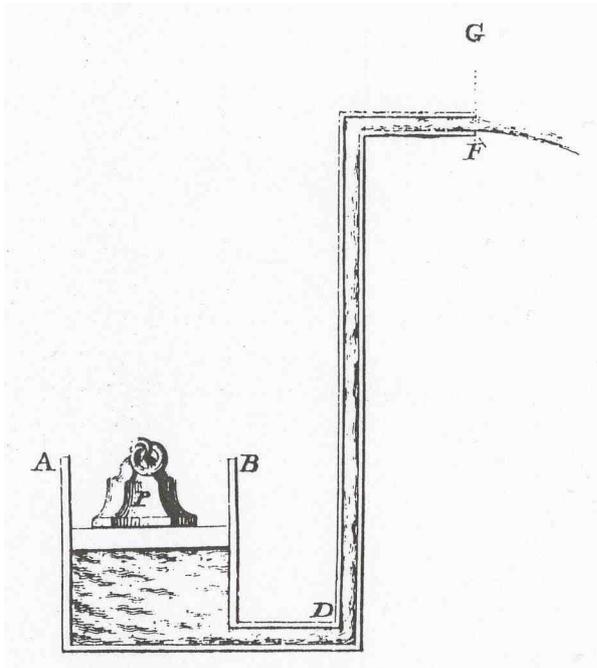

Figure 3 : Pompe à tuyau courbé

La machine devient donc compréhensible par différence avec l'idéalité choisie, cette dernière devenant ainsi une hypothèse de travail. Mais encore faut-il savoir quantifier cette perte de puissance absolue (« *dispendium potentiae absolutae* »[49]). C'est l'objet de la règle 5.

## .D. UNE PERTE DE *POTENTIA ABSOLUTA*

Considérons la Figure 3, nous dit Bernoulli dans cette règle. De l'eau en ABD est élevée plus haut en F. Par hypothèse, la vitesse moyenne sortant de F est due à la hauteur FG.

---

[48] *Ibidem* (section IX, §6). Carnot qualifie cette idée de « maxime importante » (L. CARNOT, *Principes de l'équilibre et du mouvement*, Paris, Bachelier, 1803 : p. 254, § 278).
[49] *Ibidem*: 280 (Sect. IX, §10).

Alors la perte de *potentia absoluta* est à la *potentia absoluta* totale comme FG est à la différence de hauteur entre A et G :

$$\frac{\Delta PA}{PA} = \frac{FG}{AG}$$

Pourquoi cela ? Voici la preuve avancée par le savant bâlois : si nous élargissons progressivement l'ouverture F tout en gardant le même débit, alors il arrivera un point à partir duquel on peut considérer que la vitesse sortante est insignifiante, donc proche de zéro, donc quasiment égale à la vitesse de l'eau au point G si on lui laissait la possibilité de monter jusque là. Considérons alors la puissance mouvante à l'œuvre : dans le cas de l'eau qui a une vitesse nulle en G, cette puissance mouvante $P_G$ n'est autre que le poids situé au dessus du plateau AB, poids proportionnel à la hauteur de la colonne d'eau AG dans ce cas. Mais dans notre cas, où c'est en F que l'eau acquiert une vitesse nulle, nous pouvons dire que la hauteur d'eau en cause est la différence de niveau entre le point A et le point F, et la puissance mouvante correspondante sera notée $P_F$. Donc nous avons alors deux puissances mouvantes différentes mais la vitesse de ces puissances reste identique dans les deux cas, puisque le débit à la sortie reste constant. Alors, comme la *potentia absoluta* n'est autre que la puissance mouvante par la vitesse et par le temps, et que l'on considère les mêmes durées, il vient (avec $PA_G$ la *potentia absoluta* due à la puissance mouvante $P_G$ et $PA_F$ la *potentia absoluta* due à la puissance mouvante $P_F$, et le signe $\propto$ signifiant *proportionnel à*) que la différence de *potentia absoluta* $\Delta PA$ est égal à :

$$\Delta PA = PA_G - PA_F = (P_G - P_F).v.t \propto (AG - AF).v.t = FG.v.t \quad (1)$$

La *potentia absoluta* totale étant proportionnelle quant à elle à $AG.v.t$, nous avons alors logiquement que :

$$\frac{\Delta PA}{PA} = \frac{FG}{AG}$$

La différence de *potentia absolu*ta est donc la *potentia absoluta* perdue, et elle est égale à :

$$\Delta PA = \frac{FG}{AG} \cdot PA$$

Ou, pour reprendre les notations de Bernoulli (FG=B, AF =A, et la *potentia absoluta* PA=P , *attention P ici ne représente pas le poids* P) :

$$\Delta P = \frac{B}{A+B} \cdot P$$

Curieuse manière de démontrer cela, n'est-il pas ? En effet, il lui suffirait de dire que la *potentia absoluta* est dimensionnellement identique à la force vive (c'est-à-dire P.H), et que la différence de force vive entre un fluide qui s'élève en G et un fluide qui s'élève en F est simplement alors égale à P.AG – P.AF =P. FG. Suite à quoi nous pouvons faire les rapports et assimiler cette perte de *potentia absoluta* à une perte de force vive, tellement plus habituelle dans l'œuvre de Bernoulli... Certes. Mais Bernoulli n'en fait rien.

Pourtant, quelques paragraphes plus loin dans la règle 7, Bernoulli justifiera l'existence d'une perte de la *potentia absoluta* employée dans certaines machines par le fait que : « *la montée potentielle des éléments individuels de volume du fluide* [mis en mouvement par la machine] *s'écoulant d'une cavité dans une autre à travers un orifice commun disparaît* ».[50] Mais le contexte est différent et dans ce paragraphe Bernoulli effectue un renvoi à la section précédente de son ouvrage, la section VIII dont le titre se réfère explicitement à la «*Théorie des forces vives*» et aux pertes dans le mouvement ascendant du fluide occasionnées par le passage à travers plusieurs orifices. On retrouvera une autre référence à l'utilisation des pertes de montée potentielle comme outil d'évaluation des pertes de *potentia absoluta* à la règle 8: « *La perte de potentia absoluta peut être estimée approximativement […] à partir du fait que la totalité de la montée potentielle de l'eau s'écoulant à travers la pompe peut être considérée comme produite inutilement* ».[51] Il semble donc que le mode de raisonnement de Bernoulli dans cette section dépende fortement du contexte : quand il peut raisonner exclusivement en termes de *potentia absoluta* pour évaluer le rendement d'une machine il le fait. Cela se comprend puisqu'il vient d'introduire un nouvel outil particulièrement adapté à la mesure du travail de ceux qui vont actionner ces machines, les hommes. Sa stratégie consiste alors à déterminer la part du travail fourni pour faire fonctionner la machine, dépensée en pure perte, quel que soit le type de machine utilisée. Cette stratégie s'applique donc dans les cas où la part de *potentia absoluta* dépensée inutilement peut être évaluée directement ; il n'est alors pas nécessaire de se référer à un autre domaine théorique pour déterminer le rendement d'une machine. Mais dans d'autres cas, la détermination des pertes de *potentia absoluta* nécessitera l'évaluation des déperditions intervenues dans le mouvement ascendant du fluide du fait de la production de mouvements parasites internes au fluide et dans ce cas on se référera explicitement à la *Théorie des forces vives*.

---

[50] *Ibidem* (sect. IX, § 15). Nous traduisons.
[51] *Ibidem*: 288-289 (sect. IX, § 17). Nous traduisons.

Par la suite, il démontre, par la règle 6, que si le plateau AB n'est pas exactement ajusté au piston, il se produit alors un autre type de perte de *potentia absoluta*, du fait des fuites ainsi créées. Vient après une description de la machine de Perrault, qui n'échappe pas à la règle.

Mais un nouvel argument va bientôt nous éclairer sur la manière dont Bernoulli conçoit son fameux concept. Tout ce que nous venons de dire correspondait à la première partie de la section IX, et qui traitait de machines qui éjectent de l'eau avec un certain *impetus*. Dans la seconde partie, où se trouve notre argument, Bernoulli va naturellement exposer des machines hydrauliques transportant de l'eau d'un point bas vers un point haut sans *impetus* notable. On pensera notamment aux norias, ces machines constituées de godets plongeants renversés dans la rivière, et remontant à l'endroit, servant notamment à l'irrigation.

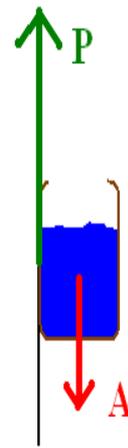

Figure 4 : Godet d'une noria sur lequel s'exercent son propre poids et la puissance mouvante.

### .E. ELARGISSEMENT DU CONCEPT DE *POTENTIA ABSOLUTA*

Bernoulli dans la deuxième partie de la section IX, se place dans les cas où les fluides sont projetés sans *impetus* notable.

#### .E.a. La *potentia absoluta* élémentaire

La règle 10[52] arrive d'entrée de jeu pour poser que dans les cas dont il est question dorénavant, la *potentia absoluta* doit être estimée par la même voie que précédemment. Or maintenant, la puissance mouvante est variable. Il convient donc d'employer une nouvelle fois le calcul différentiel et intégral, que notre homme connaît bien, pour se sortir d'affaire. Ainsi donc si un poids A monte à la hauteur *y* avec la vitesse *v* variable, et qu'il soit animé à une certaine hauteur d'une puissance mouvante *P*, l'intervalle de temps élémentaire *dt* pendant lequel le poids sera élevé de la distance *dy* sera égal à *dy/v*. Si nous multiplions celui-ci par la puissance mouvante *P* et par la vitesse *v*, nous obtenons l' « *elementum potentiae absolutae* »,[53] c'est-à-dire la *potentia absoluta* élémentaire, égale à *P.dy*. On en tire l'intégrale $\int P \cdot dy$ qui nous donne la puissance absolue totale. Cette application du calcul différentiel et intégral est essentielle dans la maturation du concept car elle permettra de penser mathématiquement les objets en mouvement non constant, en régime non stationnaire (cependant Bernoulli ne raisonne ici qu'en stationnaire). A présent, observons la figure ci contre.

---

[52] *Ibidem*: pp. 290-91.
[53] *Ibidem*.

La force résultante qui s'exerce sur le godet est *P-A*. Si nous divisons cette force par la masse du godet, et que nous multiplions le tout par *dt*, c'est-à-dire *dy/v*, nous obtenons la différentielle de la vitesse *dv*. Soit $dv = \left(\frac{P-A}{A}\right) \cdot \frac{dy}{v}$

Ou : *A.v. dv = P.dy –A.dy*

Il suffit alors d'intégrer : $\int P.dy = \frac{1}{2}Avv + A.y$ et en prenant pour les bornes de *v*, 0 , par hypothèse, et pour *y*, 0 et *a*, *a* étant la hauteur jusqu'à laquelle s'élève le godet : $\int P.dy = A.a$. Donc comme l'intégrale ci-dessus est l'expression de la *potentia absoluta* totale, celle-ci est égale à *A.a*.

### .E.b. La *potentia absoluta* ne dépend pas du chemin parcouru mais seulement de la hauteur.

C'est alors que Bernoulli fait une intéressante remarque sous forme de corollaire[54] : si au sommet de sa trajectoire le godet possède encore une vitesse résiduelle, qui lui permette de s'élever encore à une hauteur b, la *potentia absoluta* totale sera de A(a+b). Le scolie 2[55] étend ce corollaire au cas de machine élevant de l'eau non plus verticalement mais de manière inclinée. La référence à Galilée et à ses expériences sur les plans inclinés est alors évidente, et Bernoulli en vient à énoncer le scolie général[56] qui est ce que nous voulons mettre en avant. Il déclare en effet que nonobstant les frictions et les pertes de *potentia absoluta* la valeur de cette dernière, comme il a été démontré, ne dépend pas du chemin parcouru, mais simplement de la différence de hauteur entre le point le plus haut et le point le plus bas. Et il établit alors un rapprochement avec la force vive :

La *puissance absolue* a ceci *en commun* avec la force vive ou avec la descente ou remontée réelle [ascensu descensuve actuali][57]

---

[54] *Ibidem*: p. 291 (Sect. IX, § 23).
[55] *Ibidem*: p. 292 (Sect. IX, § 25).
[56] *Ibidem*.
[57] Nous soulignons.

En outre, dans la suite de la section, notamment dans la troisième partie[58] Bernoulli s'intéressera à la *potentia absoluta* qui nait suite à l'application à une machine d'un élément naturel tel que le vent. Bien que la puissance motrice ne soit plus ici les hommes ou les animaux, mais un élément naturel, Bernoulli parlera alors encore de *potentia absoluta*., puisque cette dernière se réfère à la *puissance mouvante*, la pression exercée sur la machine, quelle qu'en soit l'entité responsable. *Potentia absoluta*, et non pas travail : en effet le mot même de *travail* est réservé sous la plume de Bernoulli, aux hommes et aux animaux. Les éléments, par nature, ne sauraient *travailler*. Néanmoins la *puissance mouvante* dont ils sont capables, la pression résultante, celle dont il est question dans la première règle suscitée, peut développer une *potentia absoluta*, un travail mécanique s'exerçant par un dispositif machinique, à condition de prendre en compte une puissance mouvante continue, capable de s'auto entretenir. Maîtrisons cette subtilité, essentielle : ce ne sont pas les éléments en eux-mêmes qui sont capables de créer une *potentia absoluta*. On ne saurait en effet la rattacher proportionnellement à une mesure de leur force totale. C'est la pression, qu'ils exercent sur des machines qu'ils mettent en mouvement, qui engendre une *potentia absoluta* ensuite utilisée avec plus ou moins d'efficience par la machine. Autrement dit ce n'est que la médiation de la machine qui permet le développement de *potentia absoluta* dans le cas des éléments. Si la *potentia absoluta* est une image directe du travail des hommes, elle n'est pas image d'un hypothétique travail des éléments : non proportionnelle à la *potentia absoluta*, la force totale des éléments, quand bien même calculable en dehors de son application à une machine, ne serait pas un paramètre représentatif de l'effet produit à l'entrée de la machine. Il s'agit donc de lui substituer un concept utilisable dans un cadre productif, la *potentia absoluta*. La force totale des fluides n'est alors qu'une potentialité dont la mise en acte et en production, à jamais déficiente, passe par la machine.

---

[58] *Ibidem* : p. 305-312 (sect. IX).

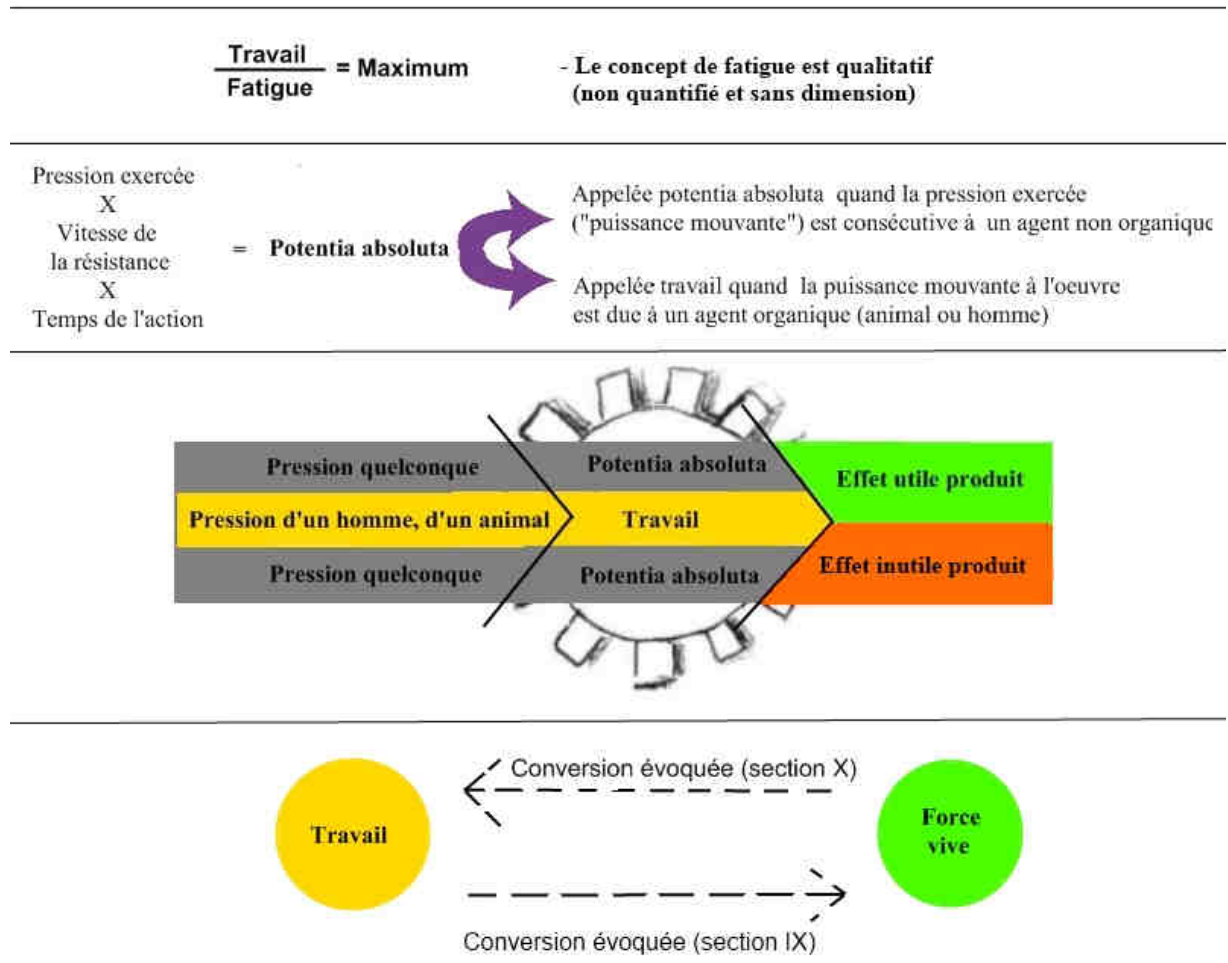

Figure 5 : Diverses caractéristiques de la fatigue, du travail, et de la *potentia absoluta*, tels que ces concepts apparaissent dans les sections IX et X de l'Hydrodynamica.

La pérennité du concept de *potentia absoluta* n'est pas simple à établir. Le mot même n'est, à notre connaissance, plus utilisée dans l'œuvre de Bernoulli, sauf une fois en 1768, mais pour désigner tout à fait autre chose, car dimensionnellement semblable à une force, et non à un travail.[59] Néanmoins on en retrouve une partie de l'acception lorsque Bernoulli parlera de travail des hommes, notamment des rameurs. Il le fera incidemment en 1742 dans

---

[59] D. BERNOULLI, *Commentation de utilissima ac commodissima directione potentiarum frictionibus mechanicis adhibendarum*, in *Die Werke von Daniel Bernoulli, Band 3, Mechanik*, éd. par D. Speiser, Basel, Boston, Stuttgart, Birkhäuser Verlag, Novi commentarii Academiae Scientiarum Imperialis Petropolitanae, vol. XIII, p. 242-256, 1768 (1769), 1987, pp. 209-218. Il fait d'ailleurs remarquer à la toute fin du texte, § 12, que la force ainsi définie est différente du « travail », car dans ce dernier on doit considérer le trajet : « Haec dum explorata habcantur, tutissime hac regula utemur, laborem quemcunque aestimandum esse ex potentia adhibita et ex motu ad directionem potentiae relato, modo simul defatigationis, quae soli incessui libero debetur, ratio habeatur. » (p. 218)

une lettre à Euler,[60] puis de manière tout à fait développé dans le prix de l'Académie Royale des Sciences de Paris qu'il remporta en 1753. La pensée de Bernoulli, cependant, subit de notables évolutions pendant les 20 ans qui séparent les premières versions de l'Hydrodynamica de ce dernier texte. Nous allons alors assister à une franche mise en contact des concepts de travail et de force vive. Ici Bernoulli considérera que le navire est une machine, ce qui est à l'époque une considération relativement neuve.[61] Cette considération lui permet d'appliquer à celui-là les concepts de travail et d'optimisation. Voyons comment.

---

[60] Lettre de Daniel Bernoulli à Leonhard Euler du 14 avril 1742 (n°XXIV) *in* P.H. FUSS (ed.), L. EULER, D. BERNOULLI, J.I. BERNOULLI, N.I. BERNOULLI, N.I. BERNOULLI &C. GOLDBACH, *Correspondance mathématique et physique de quelques célèbres géomètres du XVIIIe siècle, précédée d'une notice sur les travaux de Léonard Euler tant imprimés qu'inédits, et publiée, sous les auspices de l'Académie impériale des sciences de Saint-Pétersbourg,*, 2 vols, vol. 2, Saint-Pétersbourg, Imprimerie impériale des Sciences, 1843: 490-494. On lit, p. 493-494, dans un mélange germano-latin : « Ohne Zweifel kommen Sie mit mir in dem principio überein, dass der labor absolutus operarii pro dato tempore müsse aestimirt werden ex pressione quam exercet contra remum et velocitate quacum remum agitat in eodem puncto cui pressio applicata est. »

[61] J.-P. SERIS, *Machine et communication : du théâtre des machines à la mécanique industrielle*, Paris, Vrin, 1987: 123-157.

# L'EVOLUTION DE LA PENSEE DE BERNOULLI

Bernoulli se révèle dans ce texte beaucoup plus enclin à supprimer la distance entre travail et force vive. La volonté de dissoudre la distinction entre théorie et pratique, bien que présente précédemment, va être maintenant beaucoup plus franche, par un mouvement de mise en contact des concepts. Pourtant il semble subsister dans ce texte une sorte de hiatus, pour nous autres modernes. Si en le lisant distraitement on peut croire que l'auteur se répète, parlant tantôt de force des hommes, tantôt de travail des hommes, ou bien d'effet du travail, ou encore de force vive, en donnant l'impression au lecteur moderne que ces différences de vocable ne recouvrent qu'un souci de diversité de la langue, on comprend en y regardant de plus près que ces termes ne sont pas choisis au hasard : Bernoulli fait bien la différence entre ces termes, notamment entre travail et force vive. Et s'il parvient à l'idée que le travail puisse être une source de force vive en soi, la réciproque restera généralement invérifiée en l'absence d'une médiation machinique : la nature, seule, ne saurait produire du travail. Pour nous autres modernes il n'est pas aberrant de parler du 'travail de la pesanteur';[62] pour un homme du 18$^e$ siècle comme Bernoulli ce serait une hérésie. L'association de ces deux mots était alors dépourvue de sens. L'impression de proximité éprouvée à la lecture de notre auteur tient plutôt à notre formation, qui admet la conversion entre travail et force vive (ou plutôt énergie cinétique, suivant la traduction imparfaite du 19e s.), et à l'ambiguïté du mot travail, que ce soit dans son acception actuelle ou du 18e s., qui désigne autant le travail en tant que tel, en train de se faire, que le résultat dudit travail, autrement dit son effet.

Prenons les choses chronologiquement. Le titre de la pièce reprend l'intitulé du prix de l'Académie (*Recherches sur la manière la plus avantageuse de suppléer à l'action du Vent sur les grands Vaisseaux, soit en y appliquant les Rames, soit en y employant quelqu'autre moyen que ce puisse être*), mais y rajoute la précision suivante : *fondées sur une nouvelle Théorie de l'économie des forces et de leurs effets.*[63] Notons dès à présent cette dichotomie entre forces et effets, sur laquelle nous reviendrons plus tard. Bernoulli annonce donc dès le départ qu'il se place sur le registre de l'optimisation, au travers du mot économie, à prendre au double sens d'une harmonie générale et d'une épargne.[64]

---

[62] C'est d'ailleurs la lecture que fait par exemple de Berg du texte de Bernoulli cité plus haut.
[63] Pour étude sur l'ensemble de ce mémoire, on consultera l'excellente analyse qu'en fait Frans Cerulus dans D. BERNOULLI, *Die Werke von Daniel Bernoulli. Band 8, Technologie II*, éd. par F.A. Cerulus, Basel, Boston, Berlin, Birkhaüser, 2004: 35-72. Il y expose comment Bernoulli utilise le concept de travail dans le cadre d'une optimisation de la machine, mais passe rapidement sur le concept lui-même. Nous souhaitons ici apporter un complément sur ce que ce concept recouvre exactement dans l'esprit de Bernoulli.

Dans le cas des rames, qu'il examine en premier et durant presque la totalité de son mémoire, à l'exception des dernières pages, il s'agit, nous dit-il, de savoir si les forces employées à mouvoir le navire sont toutes utilement employés ou non, et dans l'hypothèse négative, comment atteindre à l'idéal. Ce qui nécessite d'une part une théorie sur les forces de l'homme, qui manœuvre les rames, ainsi qu'une connaissance exacte des forces requises pour propulser un navire.

La force motrice du navire, ici, est donc l'homme. Mais, c'est là tout le problème, comment la quantifier? En effet, contrairement aux éléments naturels qui fournissent en permanence de la force vive pour peu qu'on ait une source constante, les hommes peuvent éprouver la fatigue, qui les rend inapte à toute action passé un certain stade. Bernoulli le dit explicitement : la fatigue est « *la seule chose qu'il faille considérer* » (art. II, p. 4 de son mémoire). Tout le problème est donc de savoir ménager la fatigue au mieux, afin que dans le temps que dure l'action, les hommes puissent donner une action la plus grande qui soit pour une fatigue dont ils puissent récupérer en une nuit. Il reprend ici la question de la fatigue là où il l'avait laissé en 1738, dans le § 4 de la section IX de l'Hydrodynamica, pour la pousser plus loin. La fatigue en elle-même n'étant pas quantifiable, il faut donc trouver un indicateur qui lui soit proportionnel, permettant ainsi de l'estimer indirectement. Cet indicateur n'est autre que l'action dont nous venons de parler, c'est-à-dire le travail des hommes. Or, « on doit toujours estimer le travail absolu d'un homme par la pression qu'il exerce, par la vîtesse de son point d'appui & par le tems » (Ibid.), soit *P.v.t.*, supposé donc proportionnel à la fatigue, ce qui ne va pas de soi. En effet, on le sent même intuitivement, et on le constate empiriquement, les diverses manières d'obtenir un même produit P.v.t ne laisseront pas obligatoirement les hommes dans le même état de fatigue. Bernoulli prend l'exemple (p. 5) d'un homme soulevant vingt livres à trois pieds par seconde, un autre enlevant quatre livre avec une vitesse de quinze pieds par seconde, et un dernier élevant dix livres à six pieds par seconde. Si le premier pourra poursuivre son action pendant plusieurs heures, le second travail est hors des possibilités humaines, et le dernier homme se trouvera hors d'haleine au bout d'une heure.

---

[64] Le classique dictionnaire de Furetière indique à l'art Oeconomie : « Ménagement prudent que l'on fait de son bien » ; « Bel ordre et disposition des choses ». A. FURETIERE, *Dictionnaire universel, contenant généralement tous les mots françois tant vieux que modernes, et les termes de toutes les sciences et des arts*, 3 vols, La Haye & Rotterdam, A. & R. Leers, 1690.

### .A.  LA FATIGUE, OU LES QUATRE REDUCTIONS DU TRAVAIL

Bernoulli, au contraire de la section IX de l'Hydrodynamica vue précédemment, va cette fois-ci prendre le temps nécessaire afin de fonder ce principe de la proportionnalité du travail à la fatigue par quatre réductions, puisque si le travail est la seule chose qu'on puisse mesurer, c'est bien la fatigue qui est le vrai facteur limitant.

Premièrement, il s'agit de rendre la fatigue proportionnelle au temps. Ainsi, si on regarde un type de travail donné, tel que faire mouvoir une machine particulière, et bien construite, sur laquelle l'homme exercera une vitesse et une pression *constantes*, alors la seule différence proviendra du temps passé sur la machine, et, nous dit-il, « on ne sçauroit douter que dans ces circonstances les fatigues doivent être censées proportionnelles aux tems. » (§ III. P. 5) Il pose donc que la fatigue varie linéairement avec le temps, ce qui n'est pas évident puisqu'on pourrait concevoir par exemple que plus on est fatigué, plus l'effort nous coûte de fatigue à produire. Mais Bernoulli prend ici l'hypothèse qui lui semble la plus vraisemblable, car, il le dira dans le paragraphe suivant, la fatigue n'est pour lui que l'état consécutif à une dépense d'*esprits animaux*, qui chacun produisent la même action. Ainsi, appliqué à un même effort, la continuation de l'action demande le même nombre d'esprits animaux.

Secondement, il réduit le domaine d'étude, pour montrer que les fatigues sont également proportionnelles à $P.v$. En effet, nous dit-il, si on se place dans un domaine d'application limité, celui où les vitesses et les poids manipulés restent conformes à l'économie naturelle des agents agissants, alors les fatigues seront effectivement proportionnelles au produit $P.v.$ quelles que soient les valeurs de $P$ et de $v$, pourvu qu'elles soient adaptés à l'animal ou l'homme au travail.

Ces deux premiers points impliquent donc que la fatigue soit proportionnelle au travail $P.v.t$. Mais les hommes peuvent subir une infinité de travaux différents, qui mettent en œuvre une combinaison de muscles différents. Ainsi la nature des efforts mis en œuvre sera différente. « Cependant j'ai remarqué, qu'avec des fatigues égales, les hommes font constamment des effets à-peu-près égaux » (§ IV. p.6) : c'est bien là la nature de la troisième réduction. Celle-ci pouvant surprendre, il ne manquera pas de la vérifier expérimentalement dans le cours de son mémoire.

Enfin, il lui faut palier à une dernière insuffisance, qui a trait à l'inégalité entre les constitutions des hommes : on conçoit aisément qu'un homme robuste et vigoureux soit capable de plus de travail qu'un homme décharné. Peut être même le premier pourra t-il faire trois à quatre fois plus de travail pendant quelques heures de temps, tandis que le second, à la constitution beaucoup plus faible, aura besoin d'un temps beaucoup plus long pour réaliser le même travail. Mais, nous dit notre auteur, « si chacun de ces deux hommes si différents en vigueur, étoit appliqué pendant un grand nombre de jours de suite à une même sorte de travail jusqu'à se fatiguer également, je doute si leurs effets seraient forts inégaux. Cette vérité se manifeste assez clairement chez les bêtes. » (§ V. p.8)

La justification de ces deux dernières affirmations étonnantes tient à la modélisation que Bernoulli opère concernant l'origine de la fatigue. Il conçoit en effet que les muscles sont mus par des esprits animaux semblables à de petits ressorts bandés qui, en agissant, libère de la force vive, et dont le manque provoque la sensation de fatigue.[65] D'après lui, chaque humain en possèderait une même quantité, de sorte qu'il semble que « la nature ait prescrit aux animaux une certaine conservation de forces naturelles pareille à celle qu'on connoit aux forces vives produites par la pesanteur naturelle » (I, § III, p. 6). On voit que derrière cette affirmation se joue un élément théorique : la réduction au modèle de la conservation des forces vives, dans une vision élastique de la matière. Par ailleurs, il nous semble qu'il se place, dans cette démarche, dans le schème de l'égalité entre cause pleine et effet entier, utilisé entre autres par Leibniz, et refusé par D'Alembert. Nous aurons l'occasion de revenir sur ce point dans la suite.

## .B. CE QU'EST LE TRAVAIL

Le travail, après ces quatre réductions, est donc l'image exacte de la fatigue, mais ne se confond pas avec elle.[66] Qu'est-il donc ? Premièrement, on l'a vu, il s'agit formellement du produit venant de la multiplication de la résistance éprouvée par l'homme (ou l'animal), de sa vitesse et du temps de l'action, traduisant la volonté d'une *mesure*. Deuxièmement, cette mesure est celle d'un *effet*, l'effet de l'organisme humain ou animal pendant le travail, afin de réduire son activité à un phénomène quantifiable. Cet effet s'inscrit dans un cadre non-statique, et s'exprime en vainquant une *résistance*. Troisièmement il se place dans un schème

---

[65] Cette modélisation sera reprise par Lazare Carnot, qui la formulera quasiment dans les mêmes termes que Bernoulli : « Il paroît donc qu'on peut considérer quant au physique, l'animal comme un assemblage de corpuscules séparés par des ressorts plus ou moins comprimés, qui par-là recèle une certaine quantité de forces vives, et que ces ressorts en se dilatant convertissent cette force vive latente en force vive réelle. » (L. CARNOT, *Principes fondamentaux de l'équilibre et du mouvement*, Paris, 1803, pp. 246-247, § 271). Cité par O. DARRIGOL, 2001, *God, waterwheels, and molecules : Saint-Venant's anticipation of energy conservation*, « HSPS », 31, Part 2, pp. 285-353: p.310 dans une traduction anglaise.

[66] Coulomb, en citant Bernoulli, se trompe quand il évoque ce texte : « *Daniel Bernoulli […] dit que la fatigue des hommes est toujours proportionnelle à leur quantité d'action* » [dimensionnellement identique à la *potentia absoluta*] , alors que Bernoulli prend justement soin de restreindre cette proportionnalité à un domaine défini. Cf. C.-A. COULOMB, *Résultat de plusieurs expériences Destinées à déterminer la quantité d'action que les hommes peuvent fournir par leur travail journalier, suivant les différentes manières dont ils emploient leurs forces*, in BACHELIER (ed.), *Théorie des Machines Simples en ayant égard au frottement de leurs parties et à la roideur des cordages*, Paris, 1821: 257-258. Carnot commet la même erreur concernant D. Bernoulli (L. CARNOT, *Principes de l'équilibre et du mouvement*, Paris, Bachelier, 1803 : p. 252-253)

*productif*, car il est nécessaire de dépenser pour le produire le prix constitué par la fatigue, et on retire de lui un effet *optimisable*. Cette volonté d'optimisation n'est pas un jeu d'esprit : elle est à la base même du mémoire, examinant les navires comme des machines, dont on cherche qu'elles utilisent pour leur effet le maximum de la force donnée en entrée, afin d'utiliser le moins de travail qu'il soit possible au cinglage[67] des navires. Cette motivation toute économique est martelée dans tout le mémoire. Par le titre même « une nouvelle théorie de l'économie des forces & de leurs effets » ; puis « Il s'agit sans doute de connoître, avant tout autre chose, si dans l'usage des rames, les forces mouvantes sont toutes utilement employées pour mettre un navire en mouvement » (I, §I. p. 4) ; par les questions qu'il se pose : « […] quel est le plus petit nombre d'hommes possible pour fournir au dit travail pendant 8 heures par jour ; » (I, §XIV, p.14) ; et encore : « si tout l'effet est utile, on tire le meilleur parti qui soit possible du travail des hommes. » (I, §XXXVI, 39). Ainsi le mémoire de Bernoulli apparaît comme fondé par une problématique d'allocation optimale de ressource rare, dans un contexte de *production*, c'est-à-dire que l'effet produit à un certain coût, à savoir la fatigue dépensée. Une autre caractéristique, enfin, est que cette notion de travail est associé à une autre force que la pesanteur, en l'occurrence ici les forces des hommes et des autres animaux.

Ces quatre caractéristiques sont précisément celles sur lesquelles repose le concept de travail mécanique tel qu'il émerge dans la physique théorique au 19ᵉ siècle par l'entremise d'un petit groupe d'ingénieurs-savants, Coriolis en tête, et que nous avons identifié comme base de la définition de la notion de travail mécanique.[68] C'est sur cette base que nous nous permettons de qualifier les occurrences que nous avons vu, d'antécédents du concept de travail mécanique. Antécédents qui ne sont pas, bien sûr, exactement le concept tel qu'il apparaîtra au 19ᵉ siècle, mais qui entretiennent des rapports historiques identifiés par ses fondateurs mêmes.[69]

Cependant, si chez Coriolis l'effet des machines se mesurera grâce à ce nouveau concept, qui deviendra centre de son raisonnement, se transmettant de l'entrée à la sortie de la machine, Bernoulli ne raisonne pas exactement ainsi. Expliquons-nous : Bernoulli différencie nettement le travail de l'effet du travail. Et si le travail est une mesure de l'action des hommes, il ne caractérise que ce qu'on a à l'entrée de la machine. Quant à l'effet du travail des hommes, il s'agit de la quantité de force dont sont animés les objets consécutivement au travail des hommes. C'est là où nous pouvons observer à la fois une évolution sensible par rapport à son texte de 1738, ainsi que la subsistance d'un point de divergence entre force vive et travail.

---

[67] Cinglage : Chemin que fait un navire.
[68] Y. FONTENEAU, *Les antécédents du concept de travail mécanique chez Amontons, Parent et Daniel Bernoulli : de la qualité à la quantité (1699-1738)*, «Dix-Huitième Siècle», n° 41, 2009, pp. 343-368.
[69] On pourra consulter à ce propos le rapport de l'Académie des Sciences sur l'ouvrage de Coriolis, écrit par Prony, Girard et Navier, in G.-G. CORIOLIS, *Du calcul de l'effet des machines, ou considérations sur l'emploi des moteurs et sur leurs évaluation, pour servir d'introduction à l'étude spéciale des machines*, Paris, Carilian-Golury, 1829.

### .C. TRAVAIL, EFFET DU TRAVAIL ET EXIGENCE DE RENTABILITE

En effet, souvenons-nous que dans la section X de l'Hydrodynamica, dont il a été question dans la première partie de cet article, Bernoulli, comparait le travail des hommes, et l'effet obtenu par l'application à la conduite d'une machine de la force vive issue d'un pied cube de charbon ou de poudre à canon.[70] Cette relation très discrète semblait indiquer que Bernoulli admettait la conversion de la force vive en travail mécanique, pourvu qu'il existât un dispositif machinique comme intermédiaire. La conversion réciproque n'était pas évoquée explicitement[71]. C'est avec le présent texte de 1753 que celle-ci va devenir tout à fait explicite.

Nous observons ici finalement un contact entre ces concepts de travail et de force vive, beaucoup plus franc que la simple mention à laquelle nous avions droit dans la section X de l'*Hydrodynamica*, mais un contact pour ainsi dire inverse : la conversion du travail en force vive, possible sans intermédiaire. La conversion observée précédemment, de force vive en travail, subordonnée à la médiation d'un dispositif machinique, ne semble ici plus figurer : lorsqu'il traitera des machines, Bernoulli ne parlera plus que de force vive. En revanche, cette conversion force vive-travail semble se déplacer, car mentionnée comme possible dans un seul cas, celui où l'intermédiaire n'est plus la machine mais un corps organique. Quoiqu'il en soit, les éléments naturels, laissés à eux-mêmes, ne sauraient produire du travail.

Ce contact très explicite, preuve de la distance parcourue par Bernoulli depuis son précédent texte, est affirmé sans ambages à plusieurs reprises : « tout travail doit avoir son effet » (I, §XXXI, p.34) ; « nul travail ne reste sans effets » (I,§ XXXVI, p.39). Mais si le travail se mesure par le produit de la pression, de la vitesse du point d'application et du temps, « l'effet de travail peut & doit toujours être réduit à une certaine quantité de forces vives » (I, § VII, p.9). Il insiste en plusieurs occasions sur ce fait : « on mesure l'effet d'un certain travail par la quantité des forces vives qu'on a produites, soit réelles, soit potentielles » (I, § IV, p. 7) ; « il est question ici des forces vives, que l'homme produit pendant son travail » (I, §II, p.4). Certes les forces vives provenant d'un travail peuvent apparaître sous des formes diverses, néanmoins elles peuvent toutes être réduites à une certaine masse élevée à une certaine hauteur. Concrètement cela signifie par exemple que si le « travail étoit employé à donner continuellement à de nouveaux corps un certain degré de mouvement horizontal, il n'y auroit qu'à voir quelle est la hauteur verticale à la quelle ces corps pourroient s'élever avec leur vîtesse imprimée, & on aura aussitôt leur force vive sous la forme désirée » (I. § VII, p. 9), c'est-à-dire une masse multipliée par une hauteur.[72]

---

[70] D. BERNOULLI, *Die Werke von Daniel Bernoulli, Band 5, Hydrodynamik II*, éd. par G.K. Mikhailov, Basel, Boston, Berlin, Birkhaüser, 2002: 346-347 (Sect. X, § 43.). Cf. supra.
[71] Elle est mentionnée implicitement dans la section IX lorsque Bernoulli indique que pour évaluer les pertes de potentia absoluta on peut considérer que dans certains cas, règle 8 § 17, : « la totalité du potentiel ascendant est généré inutilement». Ce potentiel ascendant est généré précisément par la potentia absoluta utilisée pour faire fonctionner une pompe.

Cette connexion lui permet alors d'utiliser l'outil puissant que constitue le calcul de la force vive, pour caractériser quel doit être le travail minimal nécessaire pour faire cingler le navire. Autrement dit, il peut égaliser numériquement le travail et l'effet, posant ici de fait la nécessaire conversion de l'un en l'autre, et servant directement le but d'optimisation qu'il se donne. C'est très clair, par exemple, dans le § XII suivant (p. 12), où, se proposant de calculer l'effet du travail des hommes par la résistance que le navire vainc, et la hauteur qui générerait sa vitesse dans le cas d'une chute libre (qui est également la hauteur à laquelle le navire parviendrait s'il était projeté en l'air avec sa vitesse), il en vient à dire que l'effet du cinglage est « aussi le travail essentiellement requis » (*ibidem*), c'est-à-dire le travail minimal nécessaire au cinglage du navire s'il n'y a aucune perte. C'est très clair également dans le § XXVIII (pp. 32-33), où l'auteur, appelant $T$ le travail essentiellement requis et $E$ l'effet, conclut que dans le mouvement d'une galère, « le travail accessoire est = v/c $T$, ou bien = v/c $E$, puisque $T$, exprime un travail, qui produit son effet $E$ tout entier » (I, § XXVIII, p.33).[73] L'équivalence des mesures est ainsi posée. Mais il faut également noter une étape supplémentaire dans la formalisation du concept de travail ici, puisque, fait notable, une lettre particulière se voit assignée au concept et à sa mesure, ce que Bernoulli se gardait bien de faire en 1738 à propos de la *potentia absoluta*, se référant pour celle-ci directement à son produit.[74]

---

[72] Le fait que le travail soit une source de forces vives sera reprise par L. Carnot (L. CARNOT, *Principes de l'équilibre et du mouvement*, Paris, Bachelier, 1803 : pp. 35-36, § 56-57)
[73] Le travail accessoire correspond au travail supplémentaire nécessaire pour mouvoir la galère du au fait que le les rames ont un point d'appui mobile. Le « travail accessoire » est donc la différence entre le « travail total » et le « travail essentiellement requis ». v et c représentent respectivement la vitesse du point d'appui du mobile, et la vitesse de la galère par rapport au point d'appui.
[74] En 1699 et 1704, Guillaume Amontons et Antoine Parent, inventeurs de deux antécédents du concept de travail mécanique, ne le faisaient pas non plus.

Travail et force vive se rejoignent ainsi, numériquement équivalent, mais conceptuellement différents de par la nature des objets auxquels on les applique. Si le travail regarde du côté des hommes, la force vive est du côté des choses. En effet, le mot même de *travail* est circonscrit dans ce texte à la sphère animale et humaine. Le travail apparaît alors chez Bernoulli comme un concept entièrement animal : 'travail animal' est pour lui une sorte de tautologie. Aux éléments, on ne saurait appliquer ni le mot ni le concept. Ainsi donc, si Bernoulli a réussi à franchir la distance qui séparait travail et force vive, il ne le fait que partiellement, en ne donnant pas à l'équivalence entre cause et effet son caractère réciproque : si les hommes sont une source de forces vives, il ne s'ensuit pas que la force vive soit nécessairement produite par du travail.

Et si tout au long du mémoire le mot de travail est partout présent, revenant sans cesse dans chaque paragraphe, le contraste est saisissant avec les dernières pages, dans lesquelles le savant bâlois examine brièvement quel effet on pourrait obtenir par l'utilisation d'autres moyens, tels que « l'action du feu, d'un air condensé, d'un air échauffé, celle des vapeurs, de la poudre à canon, &c. Le vent lui-même est compris dans cette classe » (II, § XXXIV, p. 94) : le mot de travail, et le concept, disparaissent tout à fait. Alors, seule la force vive entre en jeu dans le calcul de l'effet : c'est de la force vive qu'on applique en entrée de la machine, et c'est toujours de la force vive que l'on obtient en sortie, pour exprimer l'effet. A aucun moment ceci implique que la force vive soit traitée comme le résultat d'un travail des éléments.

En effet, si les hommes sont bien une *source de force vive* (*ibidem*), l'actualisation de celle-ci dans un effet ne peut se faire que par l'intermédiaire constitué par l'organisme, humain ou animal. De même, le feu, l'air condensé, la poudre, etc., sont également une *source de force vive*, mais dont l'actualisation ne nécessite pas un travail. Toutefois, cette force vive ne peut être rendu *utile*, productive, que par l'intermédiaire d'un dispositif machine. En effet, nous dit-il, la force vive MA (c'est-à-dire exprimée par la situation équivalente d'une masse M tombant d'une hauteur A) contenue dans les choses naturelles, renferme une somme de forces motrices, qui, si elles peuvent mouvoir un dispositif, transmettent alors dans ce mouvement la force vive de ces choses naturelles.

L'effet de ces forces motrices déplacées, que Bernoulli ne fonde pas vraiment dans un concept indépendant, sont immédiatement mesurées par une force vive. Si Bernoulli pouvait développer son propos, peut être parviendrait-il à les traiter comme du travail. Car si Bernoulli affirme bien que les forces motrices déplacées ont le même *effet* que le travail des hommes puisqu'on peut les substituer à celui-ci, il ne reconditionne pas son concept de travail à ce nouvel objet que constituent les éléments impétueux.

Dans cette optique, il est significatif d'observer que le terme de *potentia absoluta* n'est plus utilisé. En effet, on ne traite plus ici, à part dans les dernières lignes, que du travail des hommes, dont la *potentia absoluta* était une image directe, certes, mais en tant qu'elle était appliquée à une machine. A présent, alors même qu'il regarde l'action d'éléments sur une machine, il ne revient pas à son terme de *potentia absoluta*, qui auparavant désignait l'effet engendré par la pression subie à l'entrée de la machine, pression non directement proportionnelle, en ce qui concerne les éléments, à leur force totale. Il n'utilise plus que le mot et concept de force vive, donnant ainsi une indication supplémentaire que, depuis 1738, il a relativement progressé dans le rapprochement entre mécanique rationnelle et science des machines. La *potentia absoluta* ne lui est désormais plus nécessaire, comme si, à l'image d'un D'Alembert, il souhaitait parvenir à une décroissance du nombre de concepts.[75]

Si les éléments ne sauraient *travailler*, ce qu'on savait déjà, la pression qu'ils exercent ne saurait donc même plus développer de la *potentia absoluta*. Les éléments se voient éloignés de la sphère de la production. Un recul paradoxal car il s'accompagne en même temps d'un rapprochement entre travail (des animaux) et force vive. Si le travail des hommes et des animaux, et donc la sphère productive, se voit ainsi en 1753 rapprochée de la mécanique rationnelle par la connexion univoque qui a lieu, les éléments, eux, se voient discrédités de toute velléité de production, et réduit à être décrit par de la force vive.

Cette discréditation est assez nettement visible lorsqu'il se prononce, dans la conclusion de son texte, sur les effets que peuvent donner les éléments, en rapport avec ceux que l'on peut espérer du travail des hommes. Dans l'utilisation de ces éléments, il existe, par nature, des pertes (en dehors des frottements) consistant dans « le mouvement qu'on donne au corps mobiles, qui servent en quelque sorte de point d'appui » (II, § XXXVII, p. 99), et dans la grande quantité des forces motrices qu'on laisse inutilement échappées faute de pouvoir les canaliser correctement. D'ailleurs, « cette dernière perte faut le plus grand défaut de toutes les machines à feu qu'on a encore imaginées » (*ibidem*). Il en conclut alors que le plus grand effet qu'on pourrait tirer des forces motrices « ne saurait jamais être assez grand pour mériter beaucoup d'attention » (*ibidem*). Ainsi, il faut « perdre toute espérance de pouvoir substituer sur les grands vaisseaux avec quelque succès considérable les forces motrices renfermées dans les choses naturelles aux travaux des hommes » (*ibidem* § XXXVI, p. 98) et donc « si nous avons donné la meilleure manière d'employer le travail des hommes pour suppléer à l'action du vent, nous avons en même tems montré de toutes les manieres praticables, qui soient possibles, celle dont on tirera le plus de profit » (*ibidem* § XXXVII, p. 99, dernière phrase du mémoire). De la sorte, il discrédite l'utilisation des éléments naturels à une quelconque prétention de rentabilité, et donc de production optimisable. Comme si, finalement, ce n'était pas dans la nature de ces éléments impétueux de pouvoir produire, contrairement à la nature de l'homme.

---

[75] Cf. à ce sujet J. VIARD, *D'Alembert et le langage scientifique: l'exemple de la force, un malentendu qui perdure*, in U. KÖLVING&I. PASSERON (ed.), *Sciences, musiques, Lumières, mélanges offerts à Anne-Marie Chouillet*, Ferney-Votaire, Centre International d'Etude du 18e siècle, 2002, pp. 93-106.

Bernoulli revient donc de ses espérances de 1738, quand il pensait encore pouvoir tirer un effet mirifique du charbon ou de la poudre à canon, en les appliquant aux mouvements des machines à feu. C'est à une singulière renonciation à laquelle nous assistons ici, car elle ne concerne pas notre auteur seul. En effet, Bernoulli semble ici vouloir refermer toute un pan de la recherche mécanique, né cinquante quatre ans plus tôt avec Amontons et son mémoire sur un moulin à feu,[76] dans lequel l'académicien démontrait qu'il était possible de substituer le feu aux hommes ou aux chevaux, de manière rentable. Revenu de ce qu'il croit être une illusion, Bernoulli semble ici réduire singulièrement le rêve technologique. L'espérance d'un rapprochement des éléments avec la sphère productive est donc battue en brèche, le travail, mot et concept, s'éloignant ainsi des éléments naturels.

## .D. LA PHILOSOPHIE SOCIALE DE BERNOULLI

Ici se révèle en filigrane un aspect que nous jugeons excessivement important, ayant trait pourrait-on dire à sa philosophie sociale : c'est parce que les humains ont un rendement supérieur que l'on doit les préférer à tout autre moyen. La logique d'optimisation justifie donc l'utilisation des humains, mais vient se greffer sur un discours de naturalisation du travail dans son acception sociale, en affirmant implicitement qu'il est dans la nature de l'homme, et non pas des éléments, de travailler. Il faut bien saisir ici que ce discours de naturalisation du travail humain et la logique d'optimisation servent d'appui l'un à l'autre : les hommes doivent travailler car ce sont eux qui ont un meilleur rendement ; les hommes ont un meilleur rendement car c'est là leur nature de travailler. Mais si l'on ne croit pas en une quelconque naturalité du travail humain, *l'aliénation de l'homme au travail est en dernière analyse fondé par l'argument de rationalisation.*

---

[76] G. AMONTONS, *Moyen de substituer commodément l'action du feu à la force des hommes et des chevaux pour mouvoir les machines*, in *MARS 1699*, Paris, Martin, Coignard fils, Guerin, 1732, pp. 112-126 (Mémoires).

Il faut également prendre conscience de l'esprit dans lequel Daniel Bernoulli travaille. Le sujet du mémoire n'est pas de substituer les machines à l'homme pour le libérer des travaux contraignants. L'ensemble composé du bateau, des rames et des humains est pris comme un système global à optimiser, froidement et mécaniquement. Il n'est ainsi jamais fait référence aux conditions de vies des rameurs, alors même que Bernoulli tire ses données d'études précédentes sur ces bateaux particuliers que sont les galères, où les conditions de vies épouvantables des chiourmes[77] ne permettaient précisément pas une efficacité optimale. Que l'on se représente l'extrême exiguïté d'un banc de galère, étudié par Zysberg, Burlet et Carrière[78] : cinq hommes devaient peiner, manger, dormir, déféquer, à l'intérieur d'un espace rectangulaire ne dépassant pas 2,30 m de long sur 1,25 de large. L'étroitesse du banc empêchait que les rameurs puissent plier les coudes, les obligeant à ramer bras raidis, dans l'incapacité donc d'utiliser les muscles des bras pour participer au mouvement ce qui, on le conçoit sans peine, ne constitue pas une manière optimale de ramer… Un poète provençal résumait ainsi les choses :

| | |
|---|---|
| La galero es nouestr'houstau | La galère est notre maison, |
| Plogue ou neve, sian a l'erto | Sommes à l'air qu'il pleuve ou neige |
| N'aven ni lansou ni cuberto […] | Nous n'avons ni drap ni couverte […] |
| Dourmen quatre ou cinq dins un ban, | Dormons à quatre ou cinq dans un banc |
| Que n'a pas tres pan de carruro, | Qui n'a pas trois pans de large, |
| Semblo tout à fet la mesuro | Et semble tout fait à la mesure |
| D'une caisso per pourta un mouert […] | D'une caisse pour mettre un mort […] |
| Fau que dins aquelle brancado […] | Il faut que dans cette brancade […] |
| Mangen et caguen tout ensen […][79] | Mangeons et chions tous ensemble […] |

Mais de cela, Daniel Bernoulli n'en a cure. Ca n'est pas son problème, pour ainsi dire. Ce qui l'intéresse est l'optimisation du système mécanique qu'il se plait à conceptualiser. Si les hommes sont rentables pour voguer à la rame, que l'on utilise donc des hommes, et peu importe leurs conditions de vie. Cette distance vis-à-vis de l'humain, Bernoulli le montrera également dans les débats sur l'inoculation.

Cet exemple nous montre qu'il faut se garder du préjugé selon lequel tous les esprits du temps des Lumières auraient eu pour but le bien être humain, et que la considération des machines étaient toujours mue par le désir charitable de soulager la pénibilité des travaux des hommes. Bernoulli fait parti de la classe dominante, et son souci est moins le bien être des acteurs, que la froide optimisation des moyens mécaniques.

---

[77] Chiourme : ensemble des rameurs d'une galère (composé en majorité de forçats, d'esclaves, aussi dénommés les Turcs, mais aussi de quelques rameurs volontaires salariés).
[78] R. BURLET, J. CARRIERE & A. ZYSBERG, *Mais comment pouvait-on ramer sur les galères du Roi-Soleil?*, «Histoire & Mesure», 1, n° 3-4, 1986, pp. 147-208.
[79] Publié dans « Lou jardin deys musos prouvençalos », Aix-en-provence, 1966, pp. 47-57, et cité par *Ibidem*: 151.

### .E. FATIGUE, TRAVAIL, FORCE VIVE : TROIS DOMAINES DIFFERENTS

Revenons maintenant sur la distinction qu'il exerce entre fatigue, travail et force vive, en reprenant la justification, vue plus haut, du fait qu'avec une fatigue égale, les hommes sont tous capables d'un effet égal. Il annonce immédiatement une théorie pouvant expliquer cet état de fait, basé sur la physiologie : d'après lui, la fatigue est causée par la perte d'esprits animaux, qui produisent le mouvement des muscles. Ainsi la sensation de fatigue ne serait qu'un déficit d'esprits animaux. Or ces esprits animaux agissent sur les muscles comme de petits ressorts bandés, qui en se débandant transmettent leur force vive, et c'est la nature d'un ressort que de produire constamment la même force vive. Et l'effet d'un travail se mesurant par la quantité de forces vives qu'on a produites, il s'ensuit qu'une fatigue égale, produit un effet égal dans toutes les espèces de travail.

Ne lisons pas distraitement ces quelques lignes, car l'épistémologie de Bernoulli se révèle ici. En effet, l'on voit que la force vive existe en quelque sorte dès l'origine, dans les éléments provoquant le mouvement. C'est bien de la force vive qui est générée par les esprits animaux considéré comme des ressorts qui se débandent. Mais cette force vive initiale n'est pas le travail, des hommes ou des animaux. Car le travail est justement ce qui permet à cette force vive issue des esprits-ressorts de pouvoir provoquer un effet, alors mesurable en force vive. Ce n'est que par l'organisme, humain ou animal, que cette potentialité peut s'exprimer en effet quantifiable. Le travail est un *état transitionnel* nécessaire entre deux formes de force vive.

Il s'agit de bien saisir cela. Même si d'un premier abord fatigue, travail et effet conçu comme force vive, peuvent apparaître comme une seule et même chose, de par leur mutuelle proportionnalité (la fatigue étant proportionnelle au travail, et tout travail produisant de la force vive), il faut bien comprendre qu'il n'y a là qu'une identité de mesure, et que la nature des objets auxquels ils s'appliquent diffèrent: la fatigue sert à caractériser l'état physiologique d'un animal ; le travail permet principalement de quantifier l'action de l'homme dans un contexte productif ; et la force vive ne caractérise que les entités naturelles et les objets. Physiologie, production et nature fondent respectivement l'essence des concepts de fatigue, travail, et force vive. Et si le travail et la fatigue sont bel et bien tous deux issus d'un organisme, celle-ci sculpte en creux l'image de celui-là. Car certes ils sont proportionnels, mais la fatigue est une dépense, puisque due au déficit d'esprit animaux, et le travail une création.

Les domaines, cependant, sont loin d'être imperméables, et nous ne les avons ainsi séparés que pour la clarté du discours. Il faut noter, en effet, que si le concept de travail se réfère de prime abord au domaine de la production, il est également physiologique, en ce qu'il s'applique aux organismes dans la production. Cependant, il peut être également uniquement physiologique, puisque Bernoulli parle du travail du cœur ou d'autres organes.[80] En outre, la force vive, bien que sa référence principale réside dans le schème conservatif du mouvement des objets naturels, est un concept à la fois naturel et productif lorsqu'il s'applique à déterminer l'effet du travail. Nous pourrions donc résumer les choses par le schéma ci-dessous.

| Domaines d'application des concepts | PHYSIOLOGIE | | | | NATURE | |
|---|---|---|---|---|---|---|
| | | | PRODUCTION | | | |
| Concepts | Fatigue | Travail | | | Force | vive |
| Entités auxquelles s'appliquent les concepts | Organisme | Cœur, organes | Hommes, animaux | | Entités mues par le travail / Machines | Eléments naturels |
| Caractéristiques | Déficit d'esprits animaux | | Rentable | | Effet du travail des organismes | Force conservative des éléments naturels |
| Mesure | Déficit de force vive, inquantifiable | P.v.t | P.v.t | | $mv^2$, Ph | $mv^2$, Ph |

Bernoulli propose en outre une ébauche de modélisation de la fatigue, qui lui permettrait de réduire ces trois concepts à deux, dans le sens où il conçoit que la fatigue n'est que le nom d'une quantité perdue de force vive, inquantifiable. Les esprits animaux, semblables à de petits ressorts bandés, recèlent une force vive potentielle, actualisée au cours de l'activité laborieuse, dont le défaut provoquera la fatigue, et qui sera en même temps la cause de l'effectuation de la tâche. Tout se passe alors *comme si* la force vive désormais réelle des esprits ressorts, non quantifiable en tant que telle, puisqu'on ne peut observer leur fonctionnement microscopique, venait à se convertir en travail grâce à l'organisme humain, dont l'*effet* est à son tour mesurable en force vive, le travail se convertissant en force vive. Nous observons alors l'idée d'une conversion de la force vive en travail, plus explicite que la mention de la section X de l'*Hydrodynamica* (cf. *supra)*. Deux fortes nuances doivent immédiatement être mises en avant : d'une part, c'est que cette conversion ne peut se faire qu'au sein d'un organisme vivant ; d'autre part, c'est que cette force vive là est inquantifiable. On n'a accès qu'au travail, ou à la force vive issue de ce travail. A ce qu'on a après, pas avant. Par rapport à la section X, on constate un point commun dans l'existence nécessaire d'un intermédiaire à la conversion, mais une différence en ce que cette médiation n'est plus portée par une machine mais par un organisme. En outre, comme nous l'avons vu précédemment, la machine dans ce texte de 1753, ne convertit pas de la force vive en travail, mais ne fait que transmettre de la force vive de l'entrée à la sortie.

---

[80] Cf. le § IV de la première partie, p. 7 : « Quant aux fonctions vitales, il seroit bien difficile de l'évaluer avec autant de justesse ; il n'y a que le travail du cœur, qu'on peut déterminer assez exactement […] » ; « […] on pourra estimer le travail journalier du cœur égal à celui d'élever 144000 livres à la hauteur d'un pied » ; puis p.8 : « J'estime le travail des muscles qui servent à la respiration plus grand » ; « […] il est à présumer que la nature a destiné les esprits animaux dans une proportion à-peu-près égale aux mouvement vitaux nécessaires & mouvement volontaires ».

Bernoulli tente donc ici une *modélisation* des esprits animaux, bien qu'elle ne soit qu'ébauchée, en voulant réduire la physiologie à la mécanique rationnelle, ce qui lui permet, à défaut de pouvoir calculer, d'introduire un modèle explicatif (cf. Figure 6). Mais ce schéma ne saurait s'appliquer lorsque la source de force vive est un agent non organique. Dans ce cas, si la médiation de la machine est nécessaire pour rendre la force vive utile, créant concomitamment et par nature une perte considérable signant l'impossibilité d'une utilisation rentable des éléments, il ne s'ensuit aucune conversion. La force vive reste force vive - Bernoulli ne parlant même plus de potentia absoluta- la seule distinction qui peut être faite résidant entre ses parties utile et inutile, la dernière, la plus importante, due à la production d'effets inutiles.

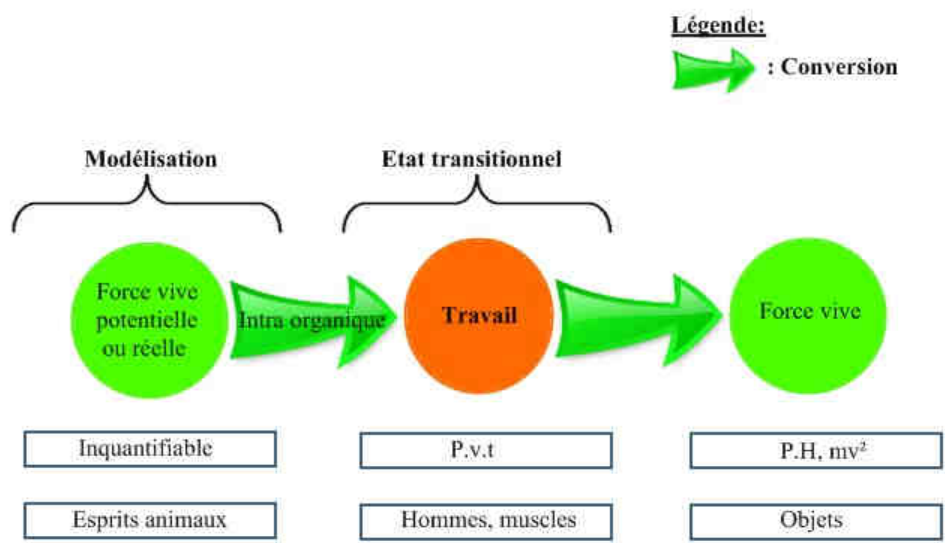

Figure 6 : La conversion du travail en force vive a lieu dans tous les cas. Il n'en va pas ainsi pour la conversion réciproque, qui n'a lieu que de manière intra organique, et qui n'est qu'une modélisation, la force vive originelle restant inquantifiable.

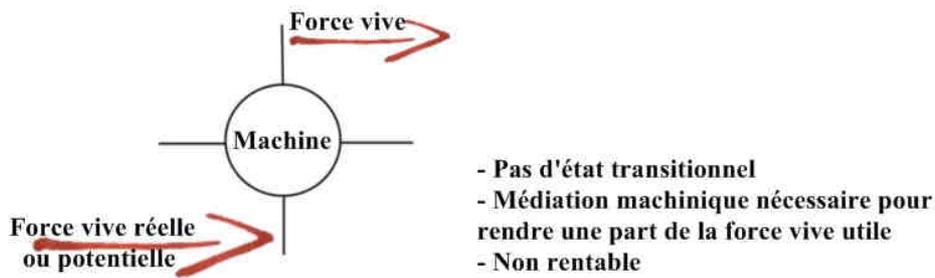

Figure 7 : Si la force vive est apportée par un agent non organique, elle est directement, bien qu'imparfaitement, utilisée par la machine et ne subit aucune espèce de conversion.

### .F.  LA POTENTIALITE MACROSCOPIQUE EST UNE FORCE VIVE REELLE MICROSCOPIQUE

Mais au-delà de ça, sa manière de concevoir la force vive peut poser question, dans le sens où elle n'est pas forcément 'réelle', ou, plus exactement, réalisée. A plusieurs reprises en effet, il considère que la force vive peut être « potentielle », et que « de même que les hommes sont une source de force vive, toutes les choses dont on pourroit se servir pour suppléer à l'action du vent renferment sous des apparences différentes une certaine quantité de forces vives. » (II, § XXXXIV, p.94), et si on évite les effets inutiles « toute la quantité de forces vives renfermées dans les choses naturelles [...] sera transmise dans les eaux poussées, et déplacées par la proue du bateau » (*ibidem*). Ce n'est pas simplement un raccourci de langage : Bernoulli semble véritablement penser que la force vive peut être contenue dans des entités n'étant pas en mouvement, comme le montre son insistance au sujet des « forces motrices renfermées dans les choses naturelles » (*ibidem* p. 98), et l'utilisation du terme de « force vive potentielle » (*ibidem* p. 97). Pour que cette force vive soit alors mesurable et que les choses naturelles réalisent leur force vive potentielle, il s'agit de l'appliquer physiquement au déplacement d'objets. C'est dans ce cadre que Bernoulli utilise la notion de force motrice : ainsi l'effet utile, calculé par lui, « résulte de la somme des forces motrices contenues dans la force vive MA » (II, § XXXV. P.95).

La force vive potentielle décrite ici désigne donc la force vive que les choses naturelles contiennent en elles, dépositaires d'une potentialité non encore effective, mais libérables à tout instant. Ceci fait sens avec d'autres écrits de lui, comme par exemple la section X de son Hydrodynamica que nous avons étudiée précédemment. En effet, nous avons vu dans la première partie de cet article que la force vive réelle des poids dont on chargeait le piston pouvait se transmettre sous forme de force vive potentielle à l'air comprimé, ainsi doté d'une capacité de stockage. En retour cette force vive potentielle pouvait inversement s'exprimer en force vive réelle. En outre, le charbon et la poudre à canon étaient dits contenir de la force vive, *latente*.[81] C'est une sorte d'énergie interne qui semblait ainsi s'exprimer.

Cette potentialité est récurrente dans l'œuvre de Bernoulli. Ainsi dans un texte de 1742,[82] il utilise également cette notion de force vive potentielle dans le cas d'une corde élastique tendue : la force vive deviendrait actuelle, si la corde était libérée de sa tension. Si ceci est sans doute insuffisant pour considérer qu'il s'agit de ce que l'on appellera plus tard une énergie potentielle, il subsiste cependant l'idée commune de la potentialité d'un effet. Il n'est donc pas étonnant de voir Bernoulli, dans son mémoire de 1753, utiliser cette notion de force vive potentielle, dès le § 7 de la première partie, lorsqu'il énonce que la mesure de l'effet d'un travail se faisait par la quantité de forces vives qu'on a produites, « soit réelles, soit potentielles » : concrètement il peut s'agir ici de ressorts qu'on a bandés. De la sorte on aura bien produit une force vive potentielle dans le double sens où elle sera contenue dans le ressort, et libérable de manière réelle.

Mais cette insistance ne peut bien se comprendre que si l'on rappelle le poids de la conception élastique de la matière sur Daniel Bernoulli, en droite ligne de son père Jean Bernoulli et de Leibniz avant lui. Pour eux, en effet, toute la matière est élastique. L'élasticité, la force élastique, est le prototype de l'énergie interne au sein de la matière.[83] Par ailleurs,

---

[81] Cf. le § B. de la première partie de cet article.
[82] Lettre de Daniel Bernoulli à Euler, citée in J. ROCHE, *What is potential energy?* , «European Journal of physics», n° 24, 2003, pp. 185-196: 189.
[83] L'élasticité est le fondement de la conservation de la force, chez ces auteurs. Leibniz l'exprime ainsi dans son Essay de dynamique : « Or, cette Elasticité des corps est nécessaire à la Nature, pour obtenir l'execution des grandes et belles loix que son Auteur infiniment sage s'est proposé, parmy lesquelles ne sont pas les moindres, ces deux Loix de la Nature que j'ay fait connoistre le premier, dont la premiere est la loy de conservation de la force absolue ou de l'action motrice dans l'univers […] et la seconde est la loy de la continuité […] Ce qui fait aussi que la nature ne souffre point de corps durs non-elastiques. » (G.W. Leibniz, *Mathematische Schriften*, vol. 6, éd. par C. Gerhardt, Hildesheim, Georg Olms Verlag, 1971: 228-229.)
Une élasticité universelle d'ailleurs refusée par Newton puis par D'Alembert qui, lui, croit aux corps durs et ne conçoit les corps élastiques que comme des corps durs munis de ressorts.(Sur la conception de D'Alembert des forces vives et les liens entre celle-ci et sa conception de l'élasticité des corps durs, cf. T.L. HANKINS, *Jean D'Alembert: Science and the enlightment*, New York, Philadelphia, London, Gordon and Breach, cop. 1990: 204-214. Pour D'Alembert, si la conservation de la force a toujours lieu pour les corps élastiques, elle est en revanche limitée, lorsqu'il s'agit des corps durs, au cas où ceux-ci agissent les uns sur les autres de façon continue, sans qu'il y ait choc : J.L.R. D'ALEMBERT &D. DIDEROT, *Encyclopédie, ou Dictionnaire raisonné des sciences, des arts et des métiers*, vol. VII, Paris, Briasson, 1757: 115 (art. FORCE).)
Jean Bernoulli, propagandiste de Leibniz, conçoit que les corps durs sont des corps élastiques pourvus d'un ressort infini, à la manière d'un « *balon rempli d'air infiniment condensé* » (Lettre de Jean Bernoulli à Dortous de Mairan du 15/06/1724, citée *in* D. Ismael Youssouf, *Les phénomènes de choc et les principes de conservation : débats historiques et processus d'apprentissage* , Thèse de doctorat, Lyon, Université Claude Bernard Lyon 1, 1999 : 77-78.) pousse encore plus loin cette conceptualisation, puisqu'il souhaite réduire tous les mouvements à ceux provoqués par la force élastique, jusqu'à faire un parallèle entre le rôle de la force élastique et celui de la pesanteur : « On peut donc considérer la chûte & l'acceleration d'un poids, comme étant causée par l'effort d'une matiere élastique, qui étenduë verticalement à l'infini, presseroit les corps de haut en bas, & les feroit pour les Bernoulli la force vive potentielle, mesure de l'élasticité de la matière, est elle-même réductible au mouvement perpétuel d'une matière subtile, et donc *in fine* à une force vive réelle microscopique.[84]

Fait remarquable, le travail des hommes lui-même semble s'inscrire dans cette conceptualisation : en effet, le travail des hommes produit de la force vive. Mais quelle est l'origine de celle-ci ? Ni plus ni moins que la fatigue des hommes, c'est-à-dire le relâchement des petits ressorts bandés constituant les esprits animaux présents dans les muscles. C'est une vision admirablement cohérente qui s'exprime ici, dans les efforts répétés de notre auteur à la suite de ses père et maître, pour réduire tout mouvement, toute production de force vive, jusqu'à travail des hommes, à l'action d'une matière élastique.

Ainsi, on peut mesurer toute la distance existant entre la force vive et sa traduction dix-neuviémiste en énergie cinétique qui, pour pratique qu'elle ait été, ne recouvrait certainement pas le même champ conceptuel. D'une part, évidemment, parce que l'énergie était un objet cognitif nouveau, associé à une vision inédite du monde non réductible à l'ancienne, ce qui est le propre des révolutions scientifiques ; mais surtout parce que l'énergie cinétique sera toujours définie comme l'énergie qu'un objet possède de par son *mouvement*, tandis que chez les Bernoulli il existe des situations dans lesquelles la force vive *prise au niveau macroscopique* peut s'interpréter en termes de potentialité, quand bien même elle puisse être interprétable en termes de mouvement au niveau microscopique. Le concept d'énergie cinétique ne saurait donc recouvrir toutes les réalités cognitives que la force vive représentait chez les divers auteurs l'utilisant, comme nous en avons ici un exemple frappant.

---

descendre selon la loy connuë de l'acceleration. » (J. Bernoulli, *Discours sur les loix de la communication du mouvement*, Paris, Jombert, 1727: 48 (chap. VII, corollaire V).)
Tout comme la pesanteur produit de la force vive, la force élastique en produit également. C'est bien cette même conception qui se retrouve à l'identique chez Daniel Bernoulli, lorsqu'il énonce dans la section X de l'Hydrodynamica : « Il doit être noté ici par avance que tout comme la descente d'un corps donné d'une hauteur donnée, de quelque manière qu'elle survienne, produit constamment la même force vive dans le corps, de même un corps élastique ou un fluide élastique, après avoir réduit son degré de tension ou de compression à un degré donné de quelque manière que ce soit, conserve en lui-même la même force vive et peut de nouveau la communiquer à d'autres corps par un échange opposé. » (D. BERNOULLI, *Die Werke von Daniel Bernoulli, Band 5, Hydrodynamik II*, éd. par G.K. Mikhailov, Basel, Boston, Berlin, Birkhäuser, 2002: 343 (sect. X, § 39).)
C'est donc bien cette conception élastique de la matière qui semble en jeu dans la notion de force vive potentielle.
[84] Cf. O. DARRIGOL, 2001, *God, waterwheels, and molecules : Saint-Venant's anticipation of energy conservation*, « HSPS », 31, Part 2: 292. J. Bernoulli énonce par exemple : « […] ces particules […] agissant avec violence contre la surface intérieure de l'endroit où elles sont renfermées, elles s'efforcent continuellement d'élargir la prison qui les retient. C'est de cet effort dont dépend la force du ressort. » (J. BERNOULLI, *Discours sur les lois de communication du mouvement*, Paris, Jombert, 1727 : 91).

# CONCLUSION : ARTICULER LA NATURE HUMAINE ET LES LOIS DE LA NATURE

Le travail chez Daniel Bernoulli est loin d'être un concept simple et analysable indépendamment, car il offre dans toutes ses occurrences une diversité de dimensions interdépendantes : mécanique, physiologique, sociale, économique. D'où ici plus encore chez d'autres auteurs l'inefficience d'une définition uniquement axée sur la pauvreté d'une expression formelle.

Le travail ne peut se comprendre que dans l'enchevêtrement de ces différentes dimensions, qui d'ailleurs se modifie un tant soit peu au cours du temps, comme on l'a vu. Toutefois, malgré la connexion explicite mais non réciproque, en dehors du cas particulier de la conversion en travail de la force vive contenue dans les muscles, entre travail et force vive que l'auteur opère en 1753, il subsiste chez lui une certaine incapacité à élargir les définitions premières de ses concepts. Chez lui, la nature ne travaille pas. Le travail, *in fine*, reste un concept humain (ou animal au sens large) tandis que la force vive s'applique aux choses naturelles.

Mais dans le même temps, il ne peut se comprendre qu'en le mettant en rapport avec la force vive ainsi qu'avec la fatigue, surtout en 1753, car le travail humain est ce qui rend possible la réalisation et la mesure des forces vives issues des esprits animaux dont la dépense provoque la fatigue. Le travail n'existe que grâce à cette dimension physiologique fatigante. Mais c'est la force vive qui permet de mesurer l'effet du travail sur les choses naturelles en 1753. C'est la force vive qui permet de mesurer dans le monde épuré de la mécanique rationnelle des éléments naturels l'incidence que la force brute des hommes a sur elle. Le travail est alors le concept nodal permettant cette mise en correspondance de l'homme et de l'univers dans le cadre d'un dispositif mécanique.

La démarche de D. Bernoulli fait alors entrer le fait biologique dans l'ère du calcul et de l'optimisation : le travail est le moyen permettant la gestion organique des hommes, en faisant passer la fatigue du domaine qualitatif au calcul quantitatif. Cette perspective est toujours envisagée par une mise en rapport avec un dispositif machinique. Il s'agit de mettre *le fait biologique* de la fatigue au centre de la conception du travail, par une interprétation mécanique. Le travail humain, en dehors de ses aspects moraux de positivité ou de négativité, est analysé comme un processus, comme un phénomène naturel, permettant la transformation de la potentialité humaine en effets mesurables sur le monde. En déterminer les lois naturelles, c'est pouvoir en régir le comportement, afin de parvenir à une optimisation. Ainsi, en 1753, le dispositif artificiel que constitue le bateau devient le point d'articulation entre les éléments naturels et la nature humaine du travailleur en action. Ici vient s'agencer la dépense organique de l'homme, et la résistance naturelle de l'eau et de la matière.

Cette démarche est tout à fait similaire à celle que Coulomb puis les ingénieurs savants du début du 19ᵉ siècle effectueront. Produire plus en se fatiguant moins, faire entrer le travail humain dans la classe des phénomènes optimisables par la connaissance des lois naturelles, afin, par suite, de rationnaliser la production.

Par suite, le concept de travail bernoullien ne pourrait se comprendre hors des cadres sociaux et économiques qu'il porte en lui au sein d'une épistémologie propre à l'auteur, qui se révèle ici animé de la marque de son temps. C'est bien le double argument de sa naturalité à l'homme et de sa nécessaire rationalisation qui fonde sa mise en avant et son utilisation. La machine qu'est le bateau n'échappe plus à cette doctrine de l'optimisation, non plus que les hommes, au travers elle. Ce n'est que parce que le travail est économiquement source de richesses et socialement valorisé, qu'il en vient à transparaître dans la mécanique des ingénieurs puis à transpirer de celle-ci à la mécanique rationnelle par toute une série de tentatives plus ou moins abouties dont celle de Bernoulli ne constitue ni la première ni la dernière, car l'élaboration du concept tel qu'il se constitue depuis Guillaume Amontons jusqu'aux ingénieurs-savants du 19ᵉ siècle, se caractérise comme une longue tentative d'arraisonnement de la mécanique des machines à la mécanique rationnelle. De la sorte, Bernoulli comme tous les autres auteurs, reprend et participe à la diffusion d'un concept et d'une valeur sociale et économique.[85]

---

[85] Sur les aspects socio-économiques du concept de travail à l'époque de Bernoulli, on peut consulter le dossier : *Penser le travail à l'époque moderne*, «Cahiers d'Histoire», n° 110, 2010.